\documentstyle[11pt,aaspp4,tighten]{article}

\received{}
\accepted{}
\journalid{}{}
\articleid{}{}

\def\VEV#1{\left\langle #1\right\rangle}
\def\epsb{\mbox{\boldmath $\epsilon$}}

\slugcomment{{\it Submitted to ApJ}}

\lefthead{Refregier \& Brown}
\righthead{Correlated noise...}

\begin{document}

\title{Effect of Correlated Noise on Source Shape Parameters and
Weak Lensing Measurements}

\author{Alexandre Refregier\altaffilmark{1}}
\affil{Department of Astrophysical Sciences, Princeton University,
Princeton, NJ 08544}

\and

\author{Scott T. Brown\altaffilmark{2}}
\affil{Department of Physics, Pupin Hall, Columbia University, New
York, NY~~10027}

\altaffiltext{1}{email: refreg@astro.princeton.edu}
\altaffiltext{2}{email: scott@astro.columbia.edu}

\begin{abstract}
The measurement of shape parameters of sources in astronomical images
is usually performed by assuming that the underlying noise is
uncorrelated. Spatial noise correlation is however present in practice
due to various observational effects and can affect source shape
parameters. This effect is particularly important for measurements of
weak gravitational lensing, for which the sought image distortions are
typically of the order of only 1\%.  We compute the effect of
correlated noise on two-dimensional gaussian fits in full
generality. The noise properties are naturally quantified by the noise
autocorrelation function (ACF), which is easily measured in
practice. We compute the resulting bias on the mean, variance and
covariance of the source parameters, and the induced correlation
between the shapes of neighboring sources. We show that these biases
are of second order in the inverse signal-to-noise ratio of the
source, and could thus be overlooked if bright stars are used to
monitor systematic distortions. Radio interferometric surveys are
particularly prone to this effect because of the long-range pixel
correlations produced by the Fourier inversion involved in their image
construction. As a concrete application, we consider the search for
weak lensing by large-scale structure with the FIRST radio survey. We
measure the noise ACF for a FIRST coadded field, and compute the
resulting ellipticity correlation function induced by the noise.  In
comparison with the weak-lensing signal expected in CDM models, the
noise correlation effect is important on small angular scales, but is
negligible for source separations greater than about $1^{\prime}$. We
also discuss how noise correlation can affect weak-lensing studies
with optical surveys.
\end{abstract}



\section{Introduction}
The measurement of source morphologies in two-dimensional images is a
fundamental problem in astronomy. The measurements of shape parameters
are usually performed while assuming that the underlying noise is
uncorrelated. However, spatial correlation of the noise is always
present to some degree, and can significantly affect the derived
parameters. In experimental situations, noise correlation can be
produced by various effects such as convolution of background light
with a beam (or point-spread function), interferometric imaging
techniques, CCD readouts, etc.

This effect is particularly important for measurements of weak
gravitational lensing. Weak lensing provides a unique opportunity to
measure the gravitational potential of massive structures along the
line-of-sight (for reviews see Schneider et al. 1992; Narayan \&
Bartelmann 1996). This technique is now routinely used to map the
potential of clusters of galaxies (see Fort \& Mellier 1994; Kaiser et
al. 1994, for reviews). Detections of the more elusive effect of
lensing by large-scale structure have been reported by Villumsen
(1995) and Schneider et al. (1997) in small optical fields. The search
for a strong detection of the effect on larger angular scales is
currently being attempted with present and upcoming wide-field CCDs in
the optical band (e.g. \markcite{ste95}Stebbins et al. 1995;
\markcite{kai96}Kaiser 1996; \markcite{ber97}Bernardeau et al. 1997),
and with the FIRST radio survey (\markcite{kam98}Kamionkowski et
al. 1998; \markcite{ref98}Refregier et al. 1998). The main challenge
comes from the fact that weak lensing induces image distortions of
only about 1\%, and thus requires high-precision measurements of
source-shape parameters. Correlated noise can produce spurious image
distortions and correlations in the shapes of neighboring sources, and
therefore must be carefully accounted for in weak lensing studies.

The effect of correlated noise is also particularly important for
radio surveys performed with interferometric arrays. In such surveys,
images are produced by Fourier inversion of the visibilities from a
set of antenna pair. As a result of the incomplete visibility
coverage, part of the noise on the image plane contains long range
correlations which can extend across an entire field. The effect of
these correlations is particularly relevant in the context of recent
large radio surveys such as FIRST (\markcite{bec95}Becker et al. 1995;
\markcite{whi96}White et al. 1996) and NVSS (Condon et
al. 1997). While the present analysis was motivated by our attempt to
measure weak lensing by large-scale structure with the FIRST radio
survey, most of the following is general and can be applied to any
wavelength and imaging technique.

\markcite{con97}Condon (1997) computed the errors in gaussian fit
parameters in the absence of noise correlation. He also gave a
semi-quantitative treatment of the noise correlation which relies on
simulations.  In the present paper, we generalize his approach to
include a general analytic treatment of the noise correlation. We show
how the spatial correlation of the noise can be naturally
characterized by the noise Auto-Correlation Function (ACF). We
explicitly derive the effect of noise correlation on source
parameters for a general two-dimensional fit and for a two-dimensional
gaussian fit. We also consider correlations induced in the parameters
of nearby sources. In particular, we focus on the ellipticity
correlation functions, which are used in searches for weak lensing by
large-scale structure. We apply our formalism to the case of the
FIRST radio survey. After measuring the noise ACF for one FIRST
coadded field, we compare the noise-induced ellipticity correlations
to those expected for weak lensing by large-scale structure. We show
that, while they are important on small angular scales, noise
correlation effects are negligible for source separations 
greater than about $1^{\prime}$.

This paper is organized as follows. In \S\ref{noise_characterization},
we define the noise ACF and show how it can be measured in
practice. In \S\ref{general_fit}, we consider a general two-dimensional
least-square fit. We derive the bias produced by the noise correlation
on the mean, variance and covariance of the fit parameters, and the
correlation of the parameters for two neighboring sources. In
\S\ref{fit_gaussian}, we apply these results to the case of
two-dimensional gaussian fits and compute the error matrix. We
explicitly derive the ellipticity correlation function induced by the
noise correlation.  In \S\ref{first}, we consider the concrete case of
the FIRST radio survey. We measure the noise ACF for a FIRST field,
compute the induced ellipticity correlation function, and compare the
latter to that expected for weak lensing. In \S\ref{optical}, we
discuss the implications of this effect for optical surveys and related
searches for weak lensing by large-scale structure. Finally,
\S\ref{conclusion} summarizes our conclusions.

\section{Characterization of the Noise}
\label{noise_characterization}
\subsection{Noise Auto-Correlation Function}
Let us consider a two-dimensional image with intensity $I({\bf x})$,
where ${\bf x}=(x_{1},x_{2})$ is the pixel position. The total
intensity can generally be decomposed into
\begin{equation}
\label{eq:i_tot}
I({\bf x})=S({\bf x})+N({\bf x}),
\end{equation}
where $S({\bf x})$ is the intensity of detected sources and N({\bf x})
is that of the noise. We take the term ``noise'' to broadly refer to
any intensity which is not associated with detected sources. Depending
on the context, it can include not only instrumental noise, but also
the background light produced by undetected sources.

We assume that, for each pixel, $N({\bf x})$
has mean and variance given by
\begin{eqnarray}
\langle N({\bf x}) \rangle & \equiv & 0 {\rm ~~~~and} \label{eq:n_mean}\\
\langle N({\bf x})^{2} \rangle & \equiv & \sigma_{N}^{2} \label{eq:n_sigma},
\end{eqnarray}
respectively. The brackets denote an ensemble average which, in
practice, is well approximated by a sky average. If necessary, the
first equation can be enforced be subtracting a constant term from
$N({\bf x})$.  The noise is generally correlated from pixel to
pixel. This is quantified by the noise ACF which we define as
\begin{equation}
\label{eq:eta_def}
\eta({\bf x}^{ab}) \equiv \langle N({\bf x}^{a})
N({\bf x}^{b}) \rangle,
\end{equation}
where ${\bf x}^{ab} \equiv {\bf x}^{b}-{\bf x}^{a}$.  

In the case of uncorrelated noise, the noise ACF becomes
\begin{equation}
\label{eq:eta_uncorr}
\eta({\bf x}) = \left\{
                   \begin{array}{ll}
                       \sigma_{N}^{2} , & \mbox{if ${\bf x}=0$} 
                                           {\rm ~~~~~~(uncorrelated)}\\
                       0 ,              & \mbox{otherwise}
		   \end{array}
                \right.
\end{equation}
In the continuous limit, i.e. in the limit of small pixel
separation $h$, this can be written as
\begin{equation}
\label{eq:eta_uncorr_cts}
\eta({\bf x}) \simeq \sigma_{N}^{2} h^{2} \delta^{(2)}({\bf x}),
{\rm ~~~~~~~~~(uncorrelated)}
\end{equation}
where $\delta^{(2)}$ is the two-dimensional Dirac-delta function.

If the noise is gaussian and homogeneous (i.e.  statistically
invariant under translations), its statistical properties are
completely characterized by $\eta({\bf x})$. Generally, however, the
noise is non-gaussian, and higher correlation functions are required
for a full description. Nevertheless, since our analysis only involves
the two-point correlation function $\eta({\bf x})$, it is also valid
in the non-gaussian case. In addition, the noise is generally not
homogeneous.  For instance, the sensitivity and beam shape can vary
accross the field-of-view and thus cause the noise properties to
depend on position. In this analysis, we assume that this effect is
small and that $\eta({\bf x})$ provides an adequate characterization
of the noise, at least in an average sense.

\subsection{Practical measurement of the noise auto-correlation function}
The noise ACF, $\eta({\bf x})$ is easily measured in practice. To do
so, discrete sources are (iteratively) removed above a given
threshold. Then, $\eta({\bf x})$ is measured by computing the average
of equation~(\ref{eq:eta_def}) over pairs of pixels separated by
$\eta({\bf x})$.

\begin{figure}
\plotone{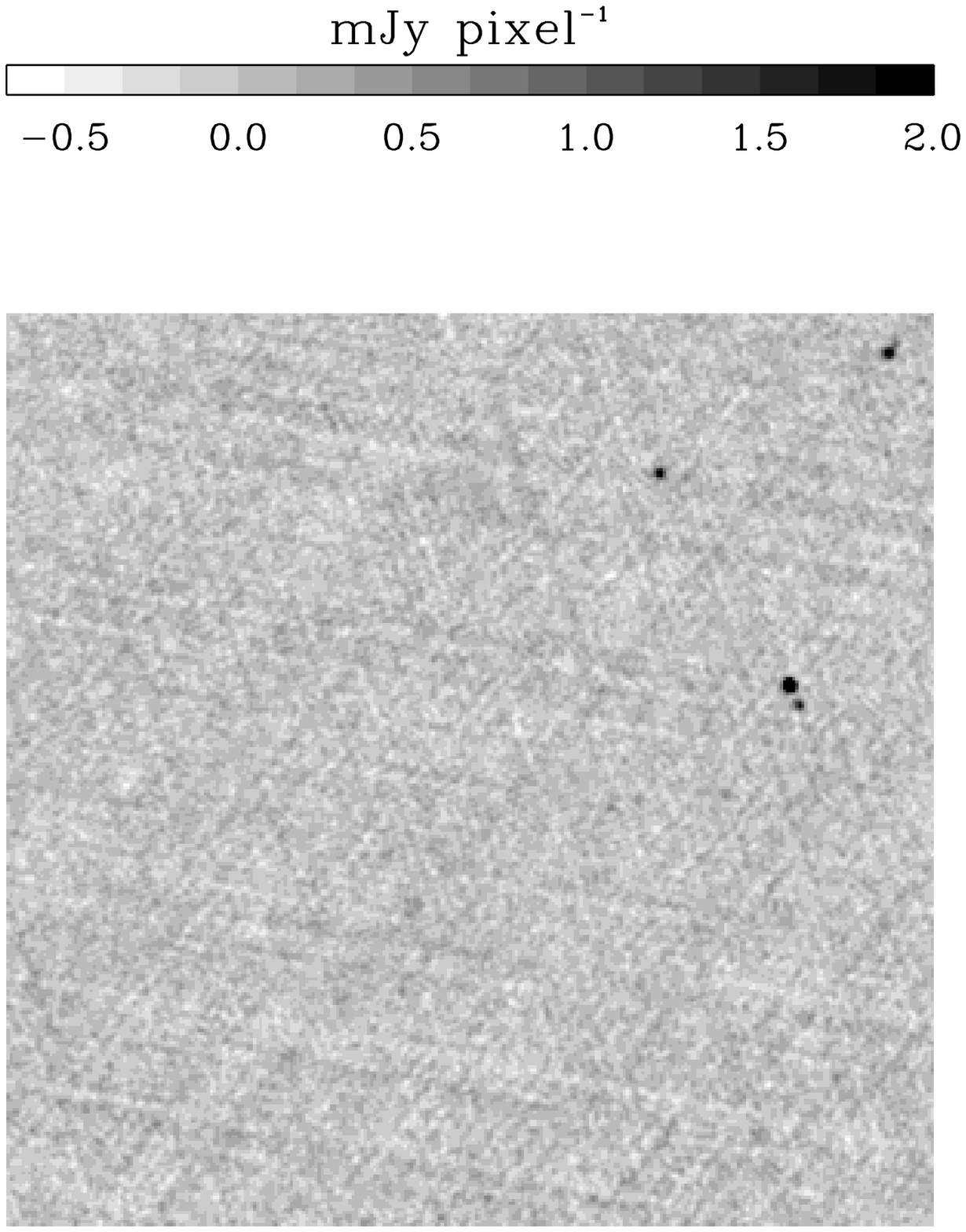}
\caption{Noise patterns for the coadded field 07210+29486E in the
FIRST survey. A $9^{\prime}\times 9^{\prime}$ region in the
$46.5^{\prime} \times 34.5^{\prime}$ of field is shown.  The contrast
for the noise was enhanced by clipping pixels with intensities higher
than 2 mJy beam$^{-1}$. The residual rms intensity is $\sigma_{N}
\simeq .15$ mJy beam$^{-1}$, for pixels at least 5.4$^{\prime \prime}$
away from pixels with intensities greater than 0.6 mJy
beam$^{-1}$. Apart from bright sources, ``stripes'' in the noise are
clearly apparent.
\label{fig:coadd}}
\end{figure}

\begin{figure}
\plotone{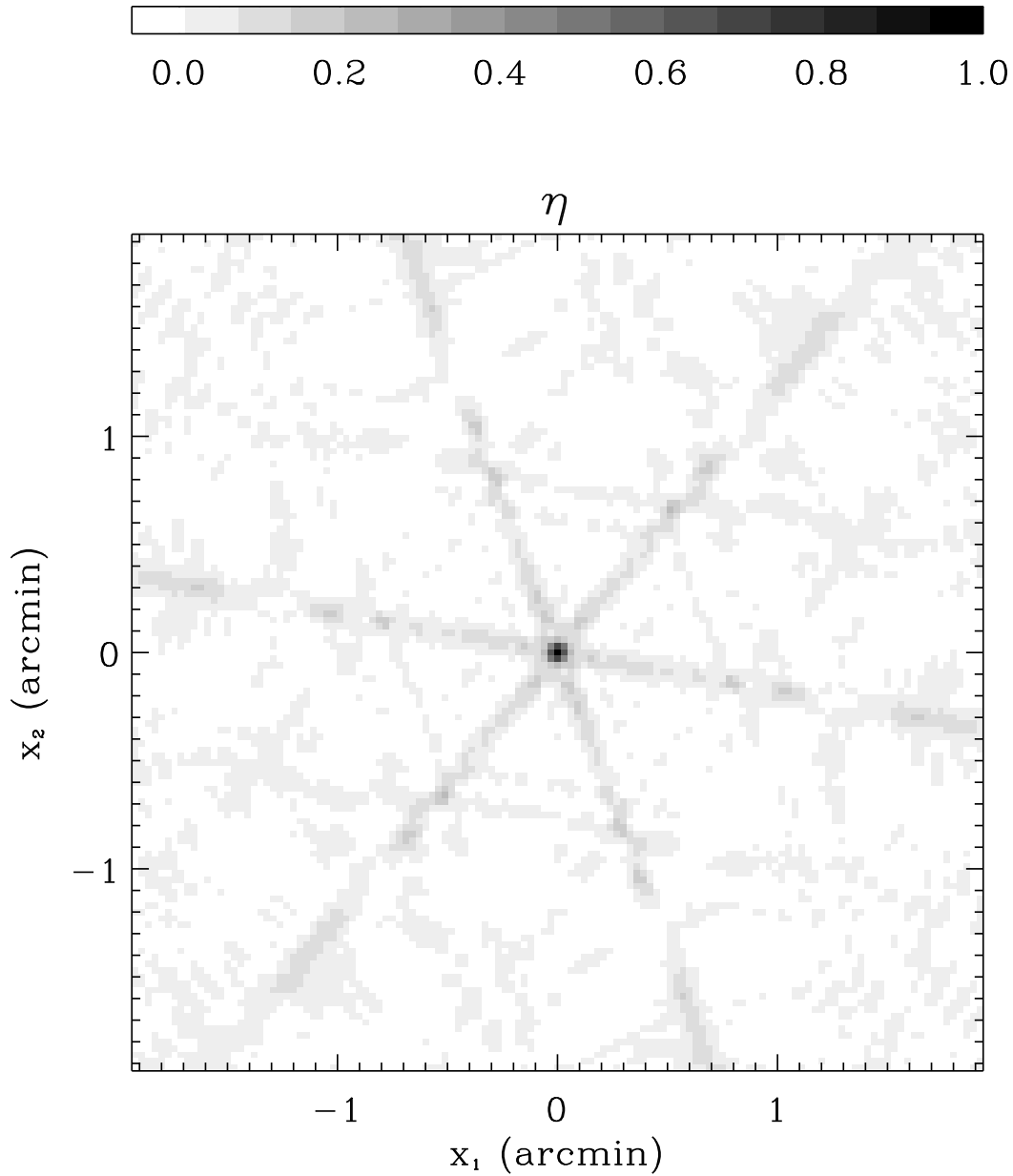}
\caption{Normalized noise auto-correlation function measured for the
coadded field 07210+29486E in the FIRST radio survey.  For clarity,
only the inner $2^{\prime} \times 2^{\prime}$ arcmin region is
shown. The correlation function was normalized to 1
at ${\bf x}=0$. For this field, $\eta(0) \equiv \sigma_{N}^{2}
\simeq 2.08 \times 10^{-8}$ mJy$^{2}$ beam$^{-2}$. For this
measurement, the SNR for $\eta(0)$ is 60.
\label{fig:eta}}
\end{figure}

Figure~\ref{fig:coadd} shows a portion of a field in the FIRST radio
survey. The contrast was enhanced to make the noise more
apparent. Stripe-like patterns in the noise are clearly visible and
indicate that the noise is correlated.  Figure~\ref{fig:eta} shows our
measurement of the noise ACF for this field. Sources were iteratively
removed by excising pixels with intensities above $4 \sigma$, and by
also removing an area $5.4^{\prime \prime}$ (1 beam width) in radius
surrounding each pixel. The measurement of $\eta({\bf x})$ was then
performed by randomly choosing pairs of pixels with separations equal
to ${\bf x}$. The long range correlations reminiscent of the VLA
antenna pattern are apparent. In \S\ref{first}, we discuss the case of
the FIRST survey in detail.

It is easy to show that, in the absence of correlation, the standard
deviation, $\sigma[\eta({\bf x})]$, of the measured ACF is
\begin{equation}
\sigma[\eta({\bf x})] = \sigma_{N}^{2}/\sqrt{N_{pairs}({\bf x})},
\end{equation}
where $N_{pairs}({\bf x})$ is the number of pixel pairs with
separation ${\bf x}$. This provides a measure of the signal-to-noise
ratio (SNR) of the ACF measurement, namely
\begin{equation}
{\rm SNR}[\eta({\bf x})] \equiv \eta({\bf x})/\sigma[\eta({\bf x})]
= \frac{\eta({\bf x})}{\eta(0)} \sqrt{N_{pairs}({\bf x})}.
\label{eq:eta_snr}
\end{equation}
In the case of figure~\ref{fig:eta}, the number of pairs was chosen to
be $N_{pairs}=3600$ for all ${\bf x}$. As a result, the SNR for
$\eta(0)$ is 60. This high value of the SNR was achieved owing to the
large number of pixels in the FIRST field.

\subsection{Noise Correlation from Beam Convolution}
\label{noise_convolution}
As Condon (1997) remarked, a natural example of noise correlation
arises when part of the noise is convolved with a beam or Point-Spread
Function (PSF). The total noise is then a sum of a convolved component
(e.g.  background light which gets smoothed by the PSF) and an
unconvolved component (e.g.  Poisson noise from the detector).  For
simplicity, we ignore the latter component and suppose that the
totality of the noise is convolved as
\begin{equation}
N({\bf x}) = \int d^{2}x^{\prime} B({\bf x}- {\bf x}^{\prime})
  \hat{N}({\bf x}^{\prime}),
\end{equation}
where $\hat{N}({\bf x})$ is the intrinsic noise which is assumed to be
uncorrelated. The convolution beam $B({\bf x})$ is assumed to be
normalized as $\int d^{2}x B({\bf x}) \equiv 1$.

As is easy to show using equation~(\ref{eq:eta_def}) and
equation~(\ref{eq:eta_uncorr_cts}) for $\hat{N}$, the resulting noise
ACF is then
\begin{equation}
\label{eq:eta_beam}
\eta({\bf x}) = \sigma_{N}^{2} \beta({\bf x}),
\end{equation}
with
\begin{equation}
\beta({\bf x}^{ab}) \equiv h^{2} \int d^{2}x B({\bf x}-{\bf x}^{a})
  B({\bf x}-{\bf x}^{b}),
\end{equation}
where, again, ${\bf x}^{ab} = {\bf x}^{b}-{\bf x}^{a}$. The function
$\beta({\bf x})$ can be thought of as the beam ACF.

\begin{figure}
\plotone{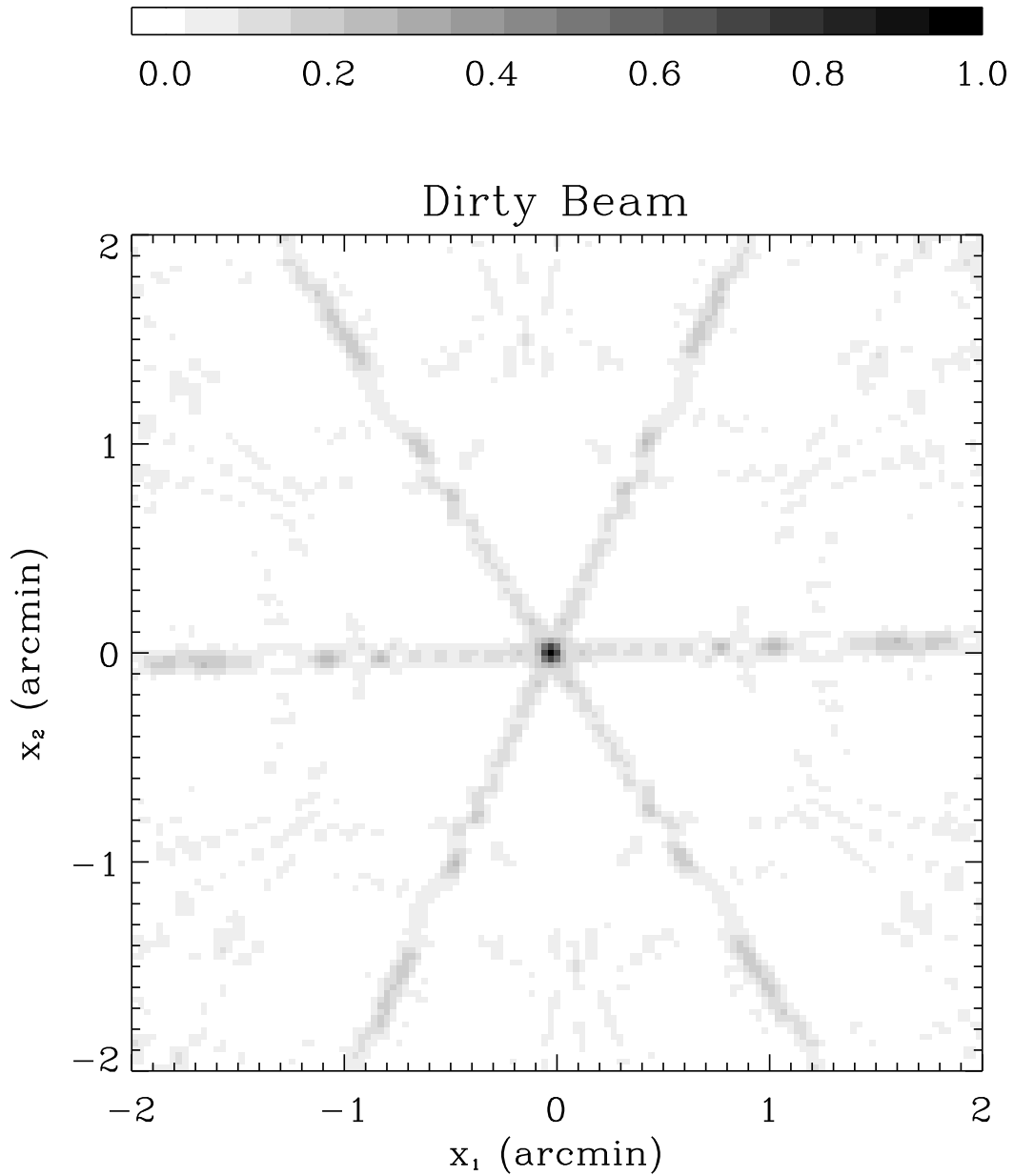}
\caption{Typical dirty beam for a grid pointing in the FIRST
survey. This particular beam was derived using AIPS for the grid
pointing 07000+50218 and was normalized to 1 at ${\bf \theta}=0$. This
pattern is typical of VLA snapshot images in which limited coverage of
the u-v plane is reflected in relatively intense sidelobes along the
projected arms of the VLA.
\label{fig:dbeam}}
\end{figure}

Figure~\ref{fig:dbeam} shows the dirty beam for a typical FIRST grid
pointing. (The dirty beam is the image of a point source placed at the
center of the field prior to CLEANing; see discussion in
\S\ref{first}).  The coadded field of figure~\ref{fig:coadd} is a
weighted sum of 13 similar grid pointings, and therefore does not have
a well defined dirty beam. It is nevertheless instructive to compare
the noise correlation function of figure~\ref{fig:eta} to this dirty
beam. Qualitatively, the resemblance is striking and reveals that most
of the noise is effectively convolved with the dirty beam.

\section{General Fit with Correlated Noise}
\label{general_fit}
There exist several methods for measuring the shape parameters of
sources. The first method consists in fitting a model (usually a
gaussian) to the two-dimensional source profile. Another possibility is
to measure the multipole moments of the source. The latter method is
usually implemented using various window functions to ensure proper
convergence (e.g. \markcite{schs95}Schneider \& Seitz 1995;
\markcite{kai95}Kaiser et al. 1995). The first method is more common
in radio interferometric imaging (e.g. AIPS software package; see
\texttt{http://www.cv.nrao.edu/aips/}), while the second is usually
used with optical images (e.g. FOCAS, SExtractor packages; see
\markcite{jar81}Jarvis \& Tyson 1981; \markcite{ber96}Bertin \& Arnouts
1996, respectively)

Here, we only consider the effect of correlated noise on model
fitting. An extension of our results to the multipole moment method is,
however, straightforward. In \S\ref{general_least_square}, we derive
the parameter corrections from the noise correlation for the case of a
general two-dimensional $\chi^2$-fit. We include corrections up to
second order in the inverse SNR of the source. In
\S\ref{general_bias}, we then compute the resulting bias in the mean
(\S\ref{general_bias}), variance, and covariance of the parameters
(\S\ref{general_covariance}).  In
\S\ref{general_correlation_function}, we compute the induced
correlation in the parameters of pairs of neighboring sources.
Finally, in \S\ref{general_functions}, we consider general functions
of the parameters. This analysis builds on the results of Condon
(1997) to include an analytical treatment of the noise correlation.

\subsection{Least Square Fit}
\label{general_least_square}
Let us consider a sector of an image consisting of a single source
superimposed on noise. We then consider a fit to the two-dimensional
source profile with a function $F({\bf x};{\bf a})$, where ${\bf
a}=(a_{1},a_{2},...)$ is a vector of parameters. In
\S\ref{fit_gaussian}, we will consider the case where $F({\bf x};{\bf
a})$ is a two dimensional gaussian with ${\bf a}$ consisting of the
gaussian normalization, peak location, and size parameters (see
Eq.~[\ref{eq:ai_gaussian}]).

We assume that the fit is ``good'', i.e., that the source profile
$S({\bf x})$ is well described by
\begin{equation}
S({\bf x}) \simeq F({\bf x};\hat{\bf a}),
\end{equation}
where $\hat{\bf a}$ is the parameter vector that would be measured
in an ideal measurement without noise. The total image intensity
(Eq.~[\ref{eq:i_tot}]) thus becomes
\begin{equation}
I({\bf x}) = F({\bf x};\hat{\bf a}) + N({\bf x}).
\end{equation}

We assume that the fit is performed using the method of least
squares, {\em without} accounting for the noise correlation.  This is
usually the case for most fitting routines (e.g. AIPS).  In other
words, the fit is performed by computing the usual functional
\begin{equation}
\chi^2({\bf a}) = \sum_{p} \frac{ \left[ I({\bf x}^{p}) - 
  F({\bf x}^{p};{\bf a}) \right]^2 }{\sigma_{N}^2},
\end{equation}
where the sum runs over all the pixels in the image and ${\bf x}^{p}$
are the position of the pixel centers. The best fit parameters ${\bf
a}$ are found by minimizing $\chi^2({\bf a})$. This minimum occurs
when $\frac{\partial \chi^2}{\partial a_{l}}({\bf a}) = 0$, that is,
for the value of ${\bf a}$ which satisfies
\begin{equation}
\label{eq:chi2_minimized}
\sum_{p} \left[ F({\bf x}^{p};\hat{\bf a}) - F({\bf x}^{p};{\bf a}) +
N({\bf x}^{p}) \right]
\frac{\partial F}{\partial a_{l}}({\bf x}^{p};\hat{\bf a}) = 0 .
\end{equation}

In the noiseless limit ($N \rightarrow 0$), an obvious, if
expected, solution is ${\bf a}=\hat{\bf a}$. Note that in the most
general case, there could exist more solutions if the system of
equation is degenerate. We will not consider this complication here
and assume that this solution is unique.
 
In the presence of noise, the solution ${\bf a}$ will deviate from
$\hat{\bf a}$. In most cases, the integrated source flux is much
larger than the integrated flux of the noise. This is quantified by
the source signal-to-noise ratio, SNR$_{s}$ which we take to be much
larger than 1. We can thus treat the noise $N({\bf x}^{p})$ as a
perturbation in equation~(\ref{eq:chi2_minimized}). For this purpose,
we rewrite $N({\bf x})$ as $\alpha N({\bf x})$, where $\alpha$ is a
dimensionless parameter of the order of $\alpha \sim {\rm
SNR}_{s}^{-1}$.  We then expand the parameter vector in powers of
$\alpha$ as ${\bf a} = \hat{\bf a}+ \alpha {\bf a}^{(1)} + \alpha^{2}
{\bf a}^{(2)} + O(\alpha^{3})$.

After inserting these expansions in
equation~(\ref{eq:chi2_minimized}), Taylor expanding, collecting terms
in powers of $\alpha$, and setting $\alpha=1$, we obtain
\begin{equation}
\label{eq:ai_general}
a_{i} = \hat{a}_{i} + D_{il} P_{l} + C_{imn} P_{m} P_{n}
   + E_{ilkm} P_{m} Q_{lk} + O({\rm SNR}_{s}^{-3}),
\end{equation}
where
\begin{eqnarray}
D_{ij} & = & (H^{-1})_{ij}, \label{eq:d_matrix}\\ 
H_{ij} & = & \sum_{p}
  \frac{\partial F}{\partial a_{i}}({\bf x}^{p};\hat{\bf a})
  \frac{\partial F}{\partial a_{j}}({\bf x}^{p};\hat{\bf a}),
  \label{eq:h_matrix}\\
B_{ijk} & = & \sum_{p}
  \frac{\partial F}{\partial a_{i}}({\bf x}^{p};\hat{\bf a})
  \frac{\partial^{2} F}{\partial a_{j}\partial a_{k}}
    ({\bf x}^{p};\hat{\bf a}), \label{eq:b_matrix}\\   
C_{imn} & = & - \left[ B_{k,rj} + \frac{1}{2} B_{r,kj} \right]
  D_{ir} D_{km} D_{jn}, \\
E_{ilkm} & = & D_{il} D_{km}, \\
P_{i} & = & \sum_{p}
  \frac{\partial F}{\partial a_{i}}({\bf x}^{p};\hat{\bf a})
  N({\bf x}^{p}), {\rm ~~~~~~and} \label{eq:p_matrix}\\
Q_{ij} & = & \sum_{p}
  \frac{\partial^{2} F}{\partial a_{i}\partial a_{j}}
    ({\bf x}^{p};\hat{\bf a})
  N({\bf x}^{p}) \label{eq:q_matrix}.
\end{eqnarray}
Unless otherwise specified, the summation convention is assumed
throughout this paper. For reasons which will become clear below, the
matrix ${\bf D}$ is called the error matrix (see Condon 1997). In the
oversampling regime, i.e. when the pixel spacing $h$ is small compared
to the source size, the sums above can be turned into integrals using
the substitution
\begin{equation}
\label{eq:cts_limit}
\sum_{p} \simeq \frac{1}{h^2} \int d^{2}x^{p}.
\end{equation}

\subsection{Bias in Fit Parameters}
\label{general_bias}
We first focus on the bias induced on the source parameters
by the noise correlation. By taking an ensemble average of
equation~(\ref{eq:ai_general}), we find the mean of the parameter
$a_{i}$ to be
\begin{equation}
\VEV{a_{i}} = \hat{a}_{i} + D_{il} \VEV{P_{l}}
  + C_{imn} \VEV{P_{m} P_{n}} + E_{ilkm} \VEV{P_{m} Q_{lk}}
  + O({\rm SNR}_{s}^{-3}).
\end{equation}
But from equation~(\ref{eq:n_mean}), we see that $\VEV{P_{l}} = 0$ so
that the first order term vanishes.  Using the definition of
$\eta({\bf x})$ (Eq.~[\ref{eq:eta_def}]) we thus find
\begin{equation}
\label{eq:ai_bias}
\VEV{a_{i}} = \hat{a}_{i} + C_{imn} V_{mn}(0) + E_{ilkm} U_{mlk}(0)
  + O({\rm SNR}_{s}^{-3}),
\end{equation}
where
\begin{eqnarray}
V_{ij}(0) & \equiv & \sum_{p} \sum_{q}
  \frac{\partial F}{\partial a_{i}}({\bf x}^{p};\hat{\bf a})
  \frac{\partial F}{\partial a_{j}}({\bf x}^{q};\hat{\bf a})
  \eta({\bf x}^{pq}), {\rm ~~~~and} \\
U_{ijk}(0) & \equiv & \sum_{p} \sum_{q}
  \frac{\partial F}{\partial a_{i}}({\bf x}^{p};\hat{\bf a})
  \frac{\partial^{2} F}{\partial a_{j} \partial a_{k}}
    ({\bf x}^{q};\hat{\bf a})
  \eta({\bf x}^{pq}).
\end{eqnarray}
The operand ``$(0)$'' was added for future convenience. In the
matrices ${\bf V}$ and ${\bf U}$, the noise ACF $\eta({\bf x})$ is
averaged over the pixel pairs weighted by parameter derivatives of the
source profile. The noise correlation therefore produces a bias
in the fit parameters of the order of SNR$_{s}^{-2}$.

Note however, that this bias does not entirely disappear in the
event of uncorrelated noise. Inserting equation~(\ref{eq:eta_uncorr})
in the previous equations, we find that, for uncorrelated noise
\begin{eqnarray}
V_{ij}(0) & = & \sigma_{N}^2 H_{ij} {\rm ~~~~~(uncorrelated)},
  \label{eq:v_uncorr}\\
U_{ijk}(0) & = & \sigma_{N}^2 B_{ijk} {\rm ~~~~(uncorrelated)}.
\end{eqnarray}
Thus, the ensemble averaged parameter becomes
\begin{equation}
\VEV{a_{i}} = \hat{a}_{i} - \frac{1}{2} \sigma_{N}^{2}
  B_{lkj} D_{li} D_{kj} + O({\rm SNR}_{s}^{-3}) {\rm ~~~~(uncorrelated)}.
\end{equation}
Therefore, even for uncorrelated noise, the $\chi^2$-fit parameters
are biased to second order in SNR$_{s}$. This is not surprising: while
the $\chi^2$-fit is always unbiased if $F({\bf x};{\bf a})$ is linear
in ${\bf a}$ (see e.g. Lupton 1993), no such guarantee exist if it is
non-linear.  It is easy to check that $\VEV{a_{i}}$ is indeed unbiased
in the linear case. However, most two-dimensional source models,
including the gaussian model, are non-linear in their parameters. The
correlation in the noise then makes this bias generally more
pronounced.

\subsection{Parameter Variance and Covariance}
\label{general_covariance}
We can also compute the covariance of the fit parameters
\begin{equation}
{\rm cov}[a_{i},a_{j}] \equiv \VEV{(a_{i}-\VEV{a_{i}})(a_{j}-\VEV{a_{j}})}.
\end{equation}
Inserting equation~(\ref{eq:ai_general}) in this definition, we
obtain
\begin{equation}
{\rm cov}[a_{i},a_{j}] = D_{il} D_{jk} V_{lk}(0) + O({\rm SNR}_{s}^{-3}).
\end{equation}
In particular, the diagonal elements yield the variance (or squared
error) of each of the fit parameters taken separately
\begin{equation}
\sigma^{2}[a_{i}] = D_{il} D_{ik} V_{lk}(0) + O({\rm SNR}_{s}^{-3}) ,
\end{equation}
where the index $i$ is not to be summed over on the right hand side.

In the case of uncorrelated noise, the covariance reduces to
\begin{equation}
{\rm cov}[a_{i},a_{j}] = \sigma_{N}^2 D_{ij}+ O({\rm SNR}_{s}^{-3})
  {\rm ~~~~~(uncorrelated)},
\end{equation}
in agreement with Condon (1997), and thereby justifying the term ``error
matrix'' applied to ${\bf D}$.  The variance obviously becomes
\begin{equation}
\sigma^{2}[a_{i}] = \sigma_{N}^2 D_{ii} + O({\rm SNR}_{s}^{-3})
  {\rm ~~~~~~~~(uncorrelated)},
\end{equation}
where, again, the index $i$ is not to be summed over.

\subsection{Parameter Correlation Function}
\label{general_correlation_function}
Let us now consider two distinct sources, $S^{1}$ and $S^{2}$, in a
single field.  We assume that the sources are sufficiently distant
from each other that the intensity of one source is negligible at the
position of the other.  If the noise in the field is uncorrelated, the
parameter fit to each of the sources would thus be independent. But if
the noise is spatially correlated, the parameters of the two sources
will be correlated. This effect is particularly relevant for weak
lensing studies in which one searches for pairwise correlations in the
ellipticities of sources.

To quantify the effect, we define the parameter correlation
function as
\begin{equation}
w^{aa}_{ij}(\hat{\bf x}^{12}) = \VEV{(a^{1}_{i}-\VEV{a^{1}_{i}})
(a^{2}_{j}-\VEV{a^{2}_{j}})},
\end{equation}
where ${\bf a}^{r}$ is the fit-parameter vector for source $S^{r}$,
and $\hat{\bf x}^{12}$ is the separation between the (noiseless)
source centroids.

Inserting equation~(\ref{eq:ai_general}) into this definition yields
\begin{equation}
w^{aa}_{ij}(\hat{\bf x}^{12}) = D_{il}^{1} D_{jk}^{2}
  V_{lk}^{12}(\hat{\bf x}^{12}) + O({\rm SNR}_{s}^{-3}) ,
\end{equation}
where,
\begin{equation}
  D_{ij}^{r} \equiv \sum_{p}
    \frac{\partial F}{\partial a_{i}}({\bf x}^{p};\hat{\bf a}^{r})
    \frac{\partial F}{\partial a_{j}}({\bf x}^{p};\hat{\bf a}^{r}),
\end{equation}
and
\begin{equation}
  V_{ij}^{12}(\hat{\bf x}^{12}) = \sum_{p} \sum_{q}
    \frac{\partial F}{\partial a_{i}}({\bf x}^{p};\hat{\bf a}^{1})
    \frac{\partial F}{\partial a_{j}}({\bf x}^{q};\hat{\bf a}^{2})
    \eta({\bf x}^{pq}).
\end{equation}
The fit-parameter correlation is therefore of order SNR$^{-2}$. 

In the absence of noise correlation (Eq.~[\ref{eq:eta_uncorr}]),
$V_{ij}^{12}(\hat{\bf x}^{12})$ becomes
\begin{equation}
\label{eq:v12_uncorr}
V_{ij}^{12}(\hat{\bf x}^{12}) = \sigma_{N}^{2} \sum_{p} 
    \frac{\partial F}{\partial a_{i}}({\bf x}^{p};\hat{\bf a}^{1})
    \frac{\partial F}{\partial a_{j}}({\bf x}^{q};\hat{\bf a}^{2}).
\end{equation}
As is the case for a gaussian fitting function, the parametric partial
derivatives, $\frac{\partial F}{\partial a_{i}}$, are often spatially
localized. In this case, the above sum vanishes, if the sources are
sufficiently distant from each other. Thus, for $|\hat{\bf x}|$ much
larger than the source sizes, and for well-behaved fitting functions
\begin{equation}
w^{aa}_{ij}(\hat{\bf x}) \simeq 0   {\rm ~~~~(uncorrelated)}.
\end{equation}
As expected, the fit parameters of the two sources are then
uncorrelated if the noise is itself spatially uncorrelated.

\subsection{Functions of the Parameters}
\label{general_functions}
We can also compute the effect of correlated noise on functions
of the parameters. Let us consider a set of such functions,
$t_{i}({\bf a})$. In \S\ref{gaussian_ellipticity}, we will
consider an application were the $t_{i}$'s are the 2 components
of the ellipticity of the source. In the presence of noise,
the value of the functions are perturbed as
(see Eq.~[\ref{eq:ai_general}])
\begin{equation}
t_{i}({\bf a}) = t_{i}(\hat{\bf a}) +
  \frac{\partial t_{i}}{\partial a_{j}}(\hat{\bf a}) D_{jk} P_{k}
  + O({\rm SNR}_{s}^{-2}).
\end{equation}
The covariance of the $t_{i}$'s is thus
\begin{eqnarray}
\label{eq:cov_tt}
{\rm cov}[t_{i},t_{j}] & \equiv &
  \VEV{ \left( t_{i}-\VEV{t_{i}} \right)
  \left( t_{j}-\VEV{t_{j}} \right)} \nonumber \\
 & = & K_{ik} K_{jl} V_{kl}(0) + O({\rm SNR}_{s}^{-3}).
\end{eqnarray}
where we have defined
\begin{equation}
K_{ij} \equiv \frac{\partial t_{i}}{\partial a_{k}}(\hat{\bf a}) D_{kj}.
\end{equation}

In the absence of noise correlation (Eq.~[\ref{eq:v_uncorr}]),
this covariance matrix becomes
\begin{equation}
{\rm cov}[t_{i},t_{j}] = \sigma_{N}^2 
  \frac{\partial t_{i}}{\partial a_{k}}(\hat{\bf a})
  \frac{\partial t_{j}}{\partial a_{l}}(\hat{\bf a})
  D_{kl} + O({\rm SNR}_{s}^{-3}), {\rm ~~~~~~(uncorrelated)}
\end{equation}
in accordance with Condon (1997).

For a pair of two distinct sources, $S^{1}$ and $S^{2}$, the
correlation function for the $t_{i}$'s is easily found to be
\begin{eqnarray}
\label{eq:wtt}
w^{tt}_{ij}(\hat{\bf x}^{12})& \equiv &
  \VEV{\left(t_{i}({\bf a}^{1})-\VEV{t_{i}({\bf a}^{1})}\right)
\left(t_{j}({\bf a}^{2})-\VEV{t_{j}({\bf a}^{2})}\right)} \nonumber \\
 & = & K_{ik}^{1} K_{jl}^{2} V_{kl}^{12}(\hat{\bf x}^{12})
       + O({\rm SNR}_{s}^{-3}).
\end{eqnarray}
where
\begin{equation}
\label{eq:k_matrix}
K_{ij}^{r} \equiv \frac{\partial t_{i}}{\partial a_{k}}(\hat{\bf a}^{r})
  D_{kj}^{r}.
\end{equation}
In the absence of noise correlation, and for distinct sources with
well-behaved fitting functions (Eq.~[\ref{eq:v12_uncorr}]), this
correlation function is simply
\begin{equation}
w^{tt}_{ij}(\hat{\bf x}) \simeq 0 {\rm ~~~~~~(uncorrelated)}
\end{equation}

\section{Gaussian Fit}
\label{fit_gaussian}
In the previous section, we derived the bias, covariance, and
correlation function of parameters for a general fit in the
presence of correlated noise. In this section, we apply these
results to the case which is of most practical interest, namely
that of a two-dimensional gaussian fit.

\subsection{Gaussian Parametrization}
As the fitting function $F({\bf x};{\bf a})$, we consider the the
following parametrization of a two-dimensional elliptical gaussian
\begin{equation}
\label{eq:gaussian_def}
G({\bf x},{\bf a}) \equiv {\cal A}
  e^{-\frac{1}{2}({\bf x}-{\bf x}^{a})^{T} {\bf A} ({\bf x}-{\bf x}^{a})},
\end{equation}
where $\cal A$ is the amplitude, ${\bf x}^a$ is the centroid vector,
${\bf A}$ is a symmetric positive-definite $2 \times 2$ matrix which
defines the shape and orientation of the gaussian, and the superscript
$^{T}$ denotes the transpose operation. We choose the parameter
vector to be 
\begin{equation}
\label{eq:ai_gaussian}
{\bf a} \equiv (x_{1}^{a}, x_{2}^{a}, {\cal A}, A_{11},
A_{12}, A_{22}).
\end{equation}
 Note that this parametrization is different from that of Condon (1997),
and was to chosen because it is easier to relate to the parameters used
in weak-lensing measurements.

It is convenient to compute the first few multipole moments $G^{(n)}$
of $G$.  With the above definition, we find
\begin{eqnarray}
G^{(0)} & \equiv & \int d^{2}x G({\bf x},{\bf a}) =
  \pi {\cal A} |A|^{-1}, \\
\frac{G_{i}^{(1)}}{G^{(0)}} & \equiv & \frac{1}{G^{(0)}}
  \int d^{2}x x_{i} G({\bf x},{\bf a}) = x_{i}^{a}, \\
\frac{G_{ij}^{(2)}}{G^{(0)}} & \equiv & \frac{1}{G^{(0)}}
  \int d^{2}x (x_{i}-x_{i}^{a}) (x_{j}-x_{j}^{a}) G({\bf x},{\bf a}) =
  (A^{-1})_{ij},
\end{eqnarray}
where $|A|$ is the determinant of ${\bf A}$. Note that 
\begin{equation}
\label{eq:j_ainverse}
{\bf J} \equiv {\bf A}^{-1} = |A|^{-1}
\left( \begin{array}{cc}
  A_{22} & -A_{12} \\
 -A_{12} &  A_{11}
\end{array} \right)
\end{equation}
is exactly equal to the normalized quadrupole moments of $G$. It
can be diagonalized as
\begin{equation}
{\bf J} = {\bf R}(-\alpha)
\left( \begin{array}{cc}
  a^{2} & 0 \\
  0 & b^{2}
\end{array} \right)
{\bf R}(-\alpha)^{T}
\end{equation}
where $a$, $b$ are the ($1 \sigma$) major and minor axes, and
$\alpha$ is the position angle measured counter-clockwise from
the positive x-axis. The rotation matrix ${\bf R}$ is
defined as 
\begin{equation}
\label{eq:rot_matrix}
{\bf R}(\varphi) \equiv \left( \begin{array}{cc}
  \cos \varphi &  \sin \varphi \\
  - \sin \varphi & \cos \varphi
\end{array} \right) . 
\end{equation}
Inverting these relations yields
\begin{eqnarray}
(a^2,b^2) & = & \frac{1}{2} \left[ J_{11}+ J_{22}
\pm \sqrt{(J_{11}-J_{22})^2+4 J_{12}^2} \right], \\
\tan 2 \alpha & = & \frac{2 J_{12}}{J_{11}-J_{22}}.
\end{eqnarray}

\subsection{Source and Noise Matrices}
We are now in a position to compute the source and noise matrices in
equations~(\ref{eq:d_matrix}-\ref{eq:q_matrix}). For this purpose,
we first compute the partial derivatives of $G$. We find,
\begin{eqnarray}
\label{eq:dgda}
\frac{\partial G}{\partial {\bf a}} & = & G  \left(
{[(x_{1}-x_{1}^{a}) A_{11} + (x_{2}-x_{2}^{a}) A_{12}],
[(x_{2}-x_{2}^{a}) A_{22} + (x_{1}-x_{1}^{a}) A_{12}], \cal A}^{-1}, \right. \nonumber \\
 & & \left.-\frac{1}{2} (x_{1}-x_{1}^{a})^{2},
-(x_{1}-x_{1}^{a})(x_{2}-x_{2}^{a}),
-\frac{1}{2} (x_{2}-x_{2}^{a})^{2} \right)
\end{eqnarray}

In the continuous limit (Eq.~[\ref{eq:cts_limit}]), the components of
the matrix ${\bf H}$ (Eq.~[\ref{eq:h_matrix}]) have a closed form, i.e.,
\begin{eqnarray}
\lefteqn{{\bf H} \simeq \frac{\pi {\cal A}^2}{h^2 \sqrt{|A|}} \times}
  \nonumber \\
 & & \left( \begin{array}{cccccc}
\frac{1}{2} A_{11} & \frac{1}{2} A_{12} & 0 & 0 & 0 & 0 \\
\frac{1}{2} A_{12} & \frac{1}{2} A_{22} & 0  & 0 & 0 & 0\\
0 & 0 & {\cal A}^{-2} &
   -\frac{1}{4} |A|^{-1} {\cal A}^{-1} A_{22} &
    \frac{1}{2} |A|^{-1} {\cal A}^{-1} A_{12} &
   -\frac{1}{4} |A|^{-1} {\cal A}^{-1} A_{11} \\
0 & 0 &-\frac{1}{4} |A|^{-1} {\cal A}^{-1} A_{22} &
   \frac{3}{16} |A|^{-2} A_{22}^2 &
  -\frac{3}{8}  |A|^{-2} A_{22} A_{12} &
   \frac{1}{16} |A|^{-2} A^2 \\
0 & 0 & \frac{1}{2} |A|^{-1} {\cal A}^{-1} A_{12} &
  -\frac{3}{8} |A|^{-2} A_{22} A_{12} &
   \frac{1}{4} |A|^{-2} A^2 &
  -\frac{3}{8} |A|^{-2} A_{11} A_{12} \\
0 & 0 & -\frac{1}{4} |A|^{-1} {\cal A}^{-1} A_{11} &
   \frac{1}{16} |A|^{-2} A^2 &
  -\frac{3}{8}  |A|^{-2} A_{11} A_{12} &
   \frac{3}{16} |A|^{-2} A_{11}^2
\end{array} \right),
\end{eqnarray}
where $A^2 \equiv (A_{11} A_{22} + 2 A_{12}^2)$. Note that the
amplitude and shape parameters mix with each other, but not with the
position parameters.

By inverting the above matrix, we find the error matrix ${\bf D}$
to be
\begin{equation}
{\bf D} \simeq \frac{2 h^{2} \sqrt{|A|}}{\pi {\cal A}^2}
\left( \begin{array}{cccccc}
|A|^{-1} A_{22} & -|A|^{-1} A_{12} & 0 & 0 & 0 & 0 \\
 -|A|^{-1} A_{12} &  |A|^{-1} A_{11} & 0 & 0 & 0 & 0 \\
0 & 0 & {\cal A}^{2} & 
  {\cal A} A_{11} & {\cal A} A_{12} & {\cal A} A_{22} \\
0 & 0 & {\cal A} A_{11} & 4 A_{11}^2 & 4 A_{11} A_{12} & 4 A_{12}^2 \\
0 & 0 & {\cal A} A_{12} & 4 A_{11} A_{12} & 2 (A_{11} A_{22} + A{12}^2) &
  4 A_{12} A_{22} \\
0 & 0 & {\cal A} A_{22} & 4 A_{12}^2 & 4 A_{22} A_{12} & 4 A_{22}^2
\end{array} \right).
\end{equation}

The ${\bf B}$ matrix of equation~(\ref{eq:b_matrix}) can be computed
in a similar way. The calculation for this $6^3$ component matrix is
however cumbersome. Since it does not enter in the parameter
correlation function on which we will now focus, we will not consider
it further.

The noise matrices ${\bf P}$ (Eq.~[\ref{eq:p_matrix}]) and ${\bf Q}$
(Eq.~[\ref{eq:q_matrix}]) can not be computed without knowledge of the
noise ACF, $\eta({\bf x})$. In practice, $\eta({\bf x})$ has a
complicated ${\bf x}$-dependence. As a result, ${\bf P}$ and ${\bf Q}$
must be computed numerically using the direct measurement of
$\eta({\bf x})$, along with the expressions for the derivatives of $G$
(Eq.~[\ref{eq:dgda}]). In the next section, we study the effect of the
noise correlation on the source ellipticity, a combination of the
source parameters which is particularly relevant for weak lensing
measurements. (By source ellipticity we mean the ellipticity of the
observed source image and not the ellipticity in the source plane, as
is common in the lensing nomenclature).

\subsection{Source Ellipticities}
\label{gaussian_ellipticity}
The shear produced by weak gravitational lensing produces distortions
in the shapes of background sources. The most direct way to detect
this effect is to look for correlations in the ellipticities of
sources. Measurements of the ellipticity involve integrals that are
more spatially extended than that of the flux and position of a
source. Consequently, we expect the ellipticity to be particularly
sensitive to the spatial correlation of the noise. Here, we compute
the magnitude of this effect on the ellipticities and on the
ellipticity correlation function.

A number of ellipticity measures have been used in weak lensing
studies (e.g. Schneider 1995; Bonnet \& Mellier 1995). Here, we will
use the ellipticity measure considered by Miralda-Escud\'{e} (1991),
Kaiser \& Squires (1993) and others, which has the advantage of being
easily related to the gaussian parameters of
equation~(\ref{eq:gaussian_def}). In this definition, the ellipticity
is defined as a two component ``vector'' $\epsb$ whose components are
given by
\begin{equation}
\epsilon_{1} = \frac{J_{11}-J_{22}}{J_{11}+J_{22}}, ~~~~~~ 
\epsilon_{2} = \frac{2 J_{12}}{J_{11}+J_{22}}.
\end{equation}
This can be rewritten in terms of the major axis, minor axis, and
position angle of the source as $\epsb = \frac{a^2-b^2}{a^2+b^2} (\cos
2 \alpha, \sin 2 \alpha)$.  The component $\epsilon_{1}$, sometimes
written as $\epsilon_{+}$, describes stretches along the $x$ and
$y$-axes, while $\epsilon_{2}$, sometimes written as
$\epsilon_{\times}$, describes stretches at $\pm 45^{\circ}$ with
respect to these axes. In a coordinate system
$(x_{1}^{\prime},x_{2}^{\prime})$ rotated by $\varphi$
counter-clockwise from the positive $x_{1}$-axis, the components of
$\epsb$ are
\begin{equation}
\label{eq:ell_rotate}
\epsilon_{i}^{\prime} = R_{ij}(2 \varphi) \epsilon_{j},
\end{equation}
where ${\bf R}$ is the rotation matrix defined in
equation~(\ref{eq:rot_matrix}). This shows that $\epsb$ is
not a true vector since it has a period of $\pi$ in $\varphi$. (In
fact, $\epsb$ can be written as a symmetric traceless $2
\times 2$ tensor).

The ellipticity can be rewritten in terms of the gaussian fit
parameters (see Eq.~[\ref{eq:j_ainverse}]) as
\begin{equation}
\epsilon_{1} = \frac{A_{22}-A_{11}}{A_{11}+A_{22}}, ~~~~
\epsilon_{2} = \frac{-2 A_{12}}{A_{11}+A_{22}}.
\end{equation}
The partial derivative matrix of $\epsb$ is thus
\begin{equation}
\frac{\partial \epsb}{\partial {\bf a}} =
2 (A_{11}+A_{22})^{-2}
\left( \begin{array}{cccccc}
0 & 0 & 0 & -A_{22} & 0 & A_{11} \\
0 & 0 & 0 & A_{12} & -(A_{11}+A_{22}) & A_{12}
\end{array} \right) .
\end{equation}
Note that the position parameters, $x_{1}^{a}$ and $x_{2}^{a}$, have
decoupled from the ellipticity matrices and can thus be dropped. There
are kept here for consistency. It is then straightforward to compute
the ${\bf K}$ matrix (Eq.~[\ref{eq:k_matrix}]). We find
\begin{equation}
{\bf K} \simeq - \frac{16 h^{2} |A|^{\frac{3}{2}}}
  {\pi {\cal A}^{2} (A_{11}+A_{22})^{2}}
\left( \begin{array}{cccccc}
0 & 0 & 0 & A_{11} & 0 & - A_{22} \\
0 & 0 & 0 & A_{12} & \frac{1}{2}(A_{11}+A_{22}) & A_{12}
\end{array} \right) .
\end{equation}
With this result and a measurement of the noise ACF $\eta({\bf x})$,
the covariance cov$[\epsilon_{i},\epsilon_{j}]$ and correlation
function $w_{ij}^{\epsilon \epsilon}({\bf x})$ of the ellipticity are
readily calculated using equations~(\ref{eq:cov_tt}) and
(\ref{eq:wtt}).

We now focus on $w_{ij}^{\epsilon \epsilon}({\bf x})$ which is of
interest for measurements of weak lensing by large-scale structure.
To be explicit, this correlation function is defined as
\begin{equation}
w_{ij}^{\epsilon \epsilon}({\bf x}^{12}) \equiv
  \VEV{ \epsilon_{i}^{1} \epsilon_{j}^{2}}
\end{equation} 
In practice, it is usually sufficient to consider the correlation of
two sources which are identical and circular. In this simplified
case, the unperturbed parameters for source $S^{1}$ and $S^{2}$ are
$\hat{\bf a}^{1}=({\cal A}, \hat{x}_{1}^{1}, \hat{x}_{2}^{1},
a^{-2},0, a^{-2})$, and $\hat{\bf a}^{2}=({\cal A}, \hat{x}_{1}^{2},
\hat{x}_{2}^{2}, a^{-2},0, a^{-2})$, where $a$ is the ($1 \sigma$)
radius of the sources, ${\cal A}$ is their amplitude, and
$\hat{x}_{i}^{r}$ gives their (unperturbed) centroid position. In this
case, the ellipticity correlation function reduces to
\begin{equation}
\label{eq:wij_circ}
w_{ij}^{\epsilon \epsilon}(\hat{x}^{12}) \simeq
  \frac{4 h^4}{\pi^{2} a^{8} {\cal A}^2}
\frac{1}{h^4} \int d^{2}x^{1} d^{2}x^{2}
X_{ij}({\bf x}^{1}-\hat{\bf x}^{1},{\bf x}^{2}-\hat{\bf x}^{2})
e^{-\frac{({\bf x}^{1}-\hat{\bf x}^{1})^2+({\bf x}^{2}-\hat{\bf x}^{2})^2}
  {2 a^{2}}}
\eta({\bf x}^{12})
+ O({\rm SNR}_{s}^{-3}),
\end{equation}
where
\begin{equation}
{\bf X}(\mbox{\boldmath $\xi$}^{1},\mbox{\boldmath $\xi$}^{2}) \equiv
\left( \begin{array}{cc}
[(\xi_{1}^{1})^{2}-(\xi_{2}^{1})^{2}] [(\xi_{1}^{2})^{2}-(\xi_{2}^{2})^{2}] &
2 \xi_{1}^{2} \xi_{2}^{2} [(\xi_{1}^{1})^{2}-(\xi_{2}^{1})^{2}] \\
2 \xi_{1}^{1} \xi_{2}^{1} [(\xi_{1}^{2})^{2}-(\xi_{2}^{2})^{2}] &
4 \xi_{1}^{1} \xi_{2}^{1} \xi_{1}^{2} \xi_{2}^{2} 
\end{array} \right).
\end{equation}

It is also convenient to consider the ellipticities measured with
respect to axes parallel and perpendicular to ${\bf x}^{12}$, the
vector connecting two sources (\markcite{kam98} Kamionkowski et
al. 1998). Let us write this vector in polar coordinates as ${\bf
x}^{12} = (\theta^{12},\varphi^{12})$, where $\theta^{12}$ is the norm
of the vector and $\varphi^{12}$ the angle it substands with the
$x_{1}$-axis.  From equation~(\ref{eq:ell_rotate}), the components of
the ellipticities rotated into this coordinate system are
$\epsilon_{i}^{r} \equiv R_{ij}(2 \varphi^{12}) \epsilon_{j}$.  The
correlation function of the rotated ellipticities is then
\begin{equation}
w_{ij}^{rr}(\theta^{12},\varphi^{12}) \equiv 
\VEV{ \epsilon_{i}^{r,1} \epsilon_{j}^{r,2} } = 
R_{ik}(2 \varphi^{12}) R_{jl}(2 \varphi^{12}) 
  w_{kl}^{\epsilon \epsilon}(\theta^{12},\varphi^{12}).
\label{eq:wijrr}
\end{equation}
Following the notation of \markcite{kai92}Kaiser (1992), we identify
$w_{11}^{rr}({\bf x}) \equiv C_{1}({\bf x})$, and $w_{22}^{rr}({\bf
x}) \equiv C_{2}({\bf x})$. We can also define a third independent
correlation function $w_{12}^{rr}({\bf x}) \equiv w_{21}^{rr}({\bf x})
\equiv C_{3}({\bf x})$, which should vanish if the noise is invariant
under parity conservation.

These correlation functions can then be averaged over $\varphi^{12}$
by defining
\begin{equation}
\bar{w}_{ij}^{rr}(\theta) \equiv \frac{1}{\pi} \int_{0}^{\pi}
d\varphi w_{ij}^{rr}(\theta,\varphi).
\label{eq:wijrr_bar}
\end{equation}
In the above notation, we simply write the azimuthally averaged
correlation functions as $\bar{w}_{11}^{rr}(\theta) \equiv C_{1}(\theta)$,
$\bar{w}_{22}^{rr}(\theta) \equiv C_{2}(\theta)$ and
$\bar{w}_{12}^{rr}(\theta) \equiv \bar{w}_{21}^{rr}(\theta) \equiv
C_{3}(\theta)$.  In the context of weak lensing studies,
$C_{1}(\theta)$ and $C_{2}(\theta)$ can be directly related to the
power spectrum of density perturbations along the line of sight (see
e.g. Kaiser 1992).

\begin{figure}
\plotone{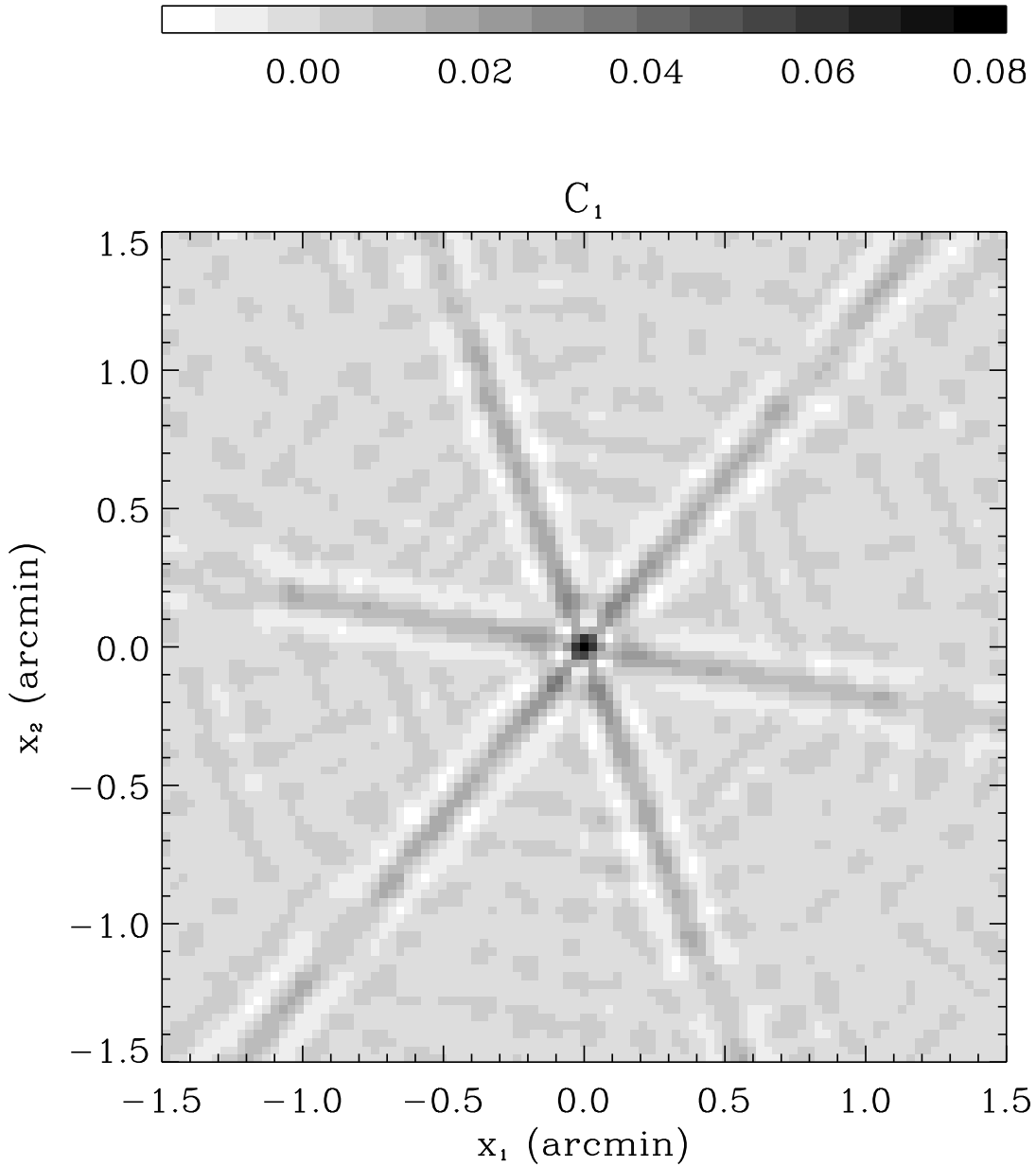}
\caption{Ellipticity correlation functions for the
coadded field 07210+29486E. Each correlation function
$C_{1}$, $C_{2}$ and $C_{3}$ is plotted as a function of the
separation vector ${\bf x}$ in panels a, b, and c, respectively.
For each source pair member, source signal-to-noise ratio was set to
SNR$_{s}=5$, while the $1 \sigma$ source size was set to
$a=2.45^{\prime \prime}$. This corresponds to FWHM convolved and
deconvolved diameters of $5.76^{\prime \prime}$ and $2^{\prime \prime}$,
respectively. The clean beam for the FIRST survey has a FWHM diameter of
$5.4^{\prime \prime}$. \label{fig:ci_th}}
\end{figure}

Figure~\ref{fig:ci_th} shows the rotated correlation functions for the
FIRST coadded field of figures~\ref{fig:coadd} and
\ref{fig:eta}. A source size of $a=2.45^{\prime \prime}$ and
signal-to-noise ratio of SNR$_{s}=5$ were
chosen. Figures~\ref{fig:ci_a}, \ref{fig:ci_err}, and
\ref{fig:ci_theo} show the corresponding azimuthally averaged
correlation functions. In the next section, we will describe these
results in detail and discuss their implications for weak lensing
measurements.

\begin{figure}
\figurenum{4b}
\plotone{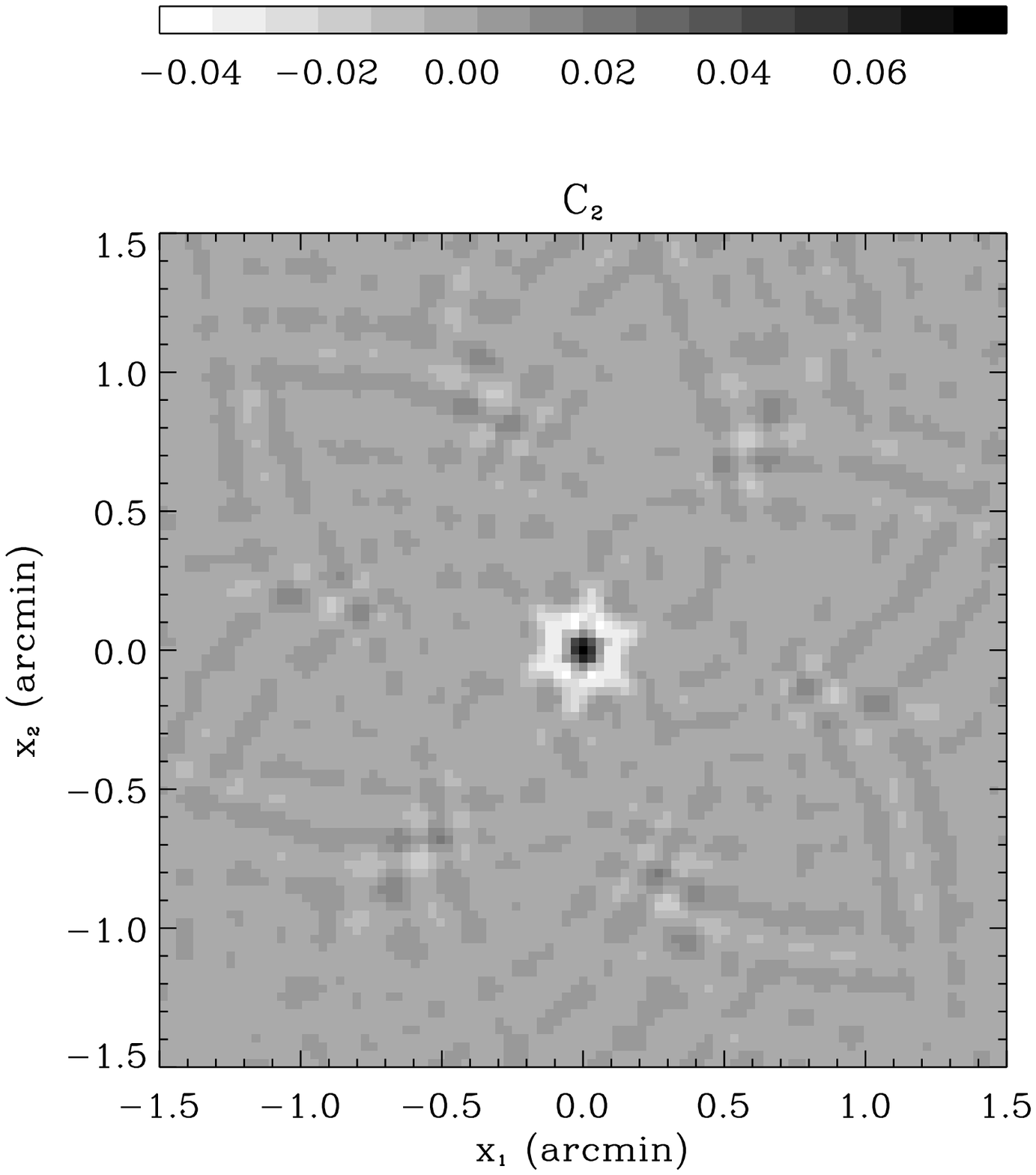}
\caption{[See caption above]}
\end{figure}

\begin{figure}
\figurenum{4c}
\plotone{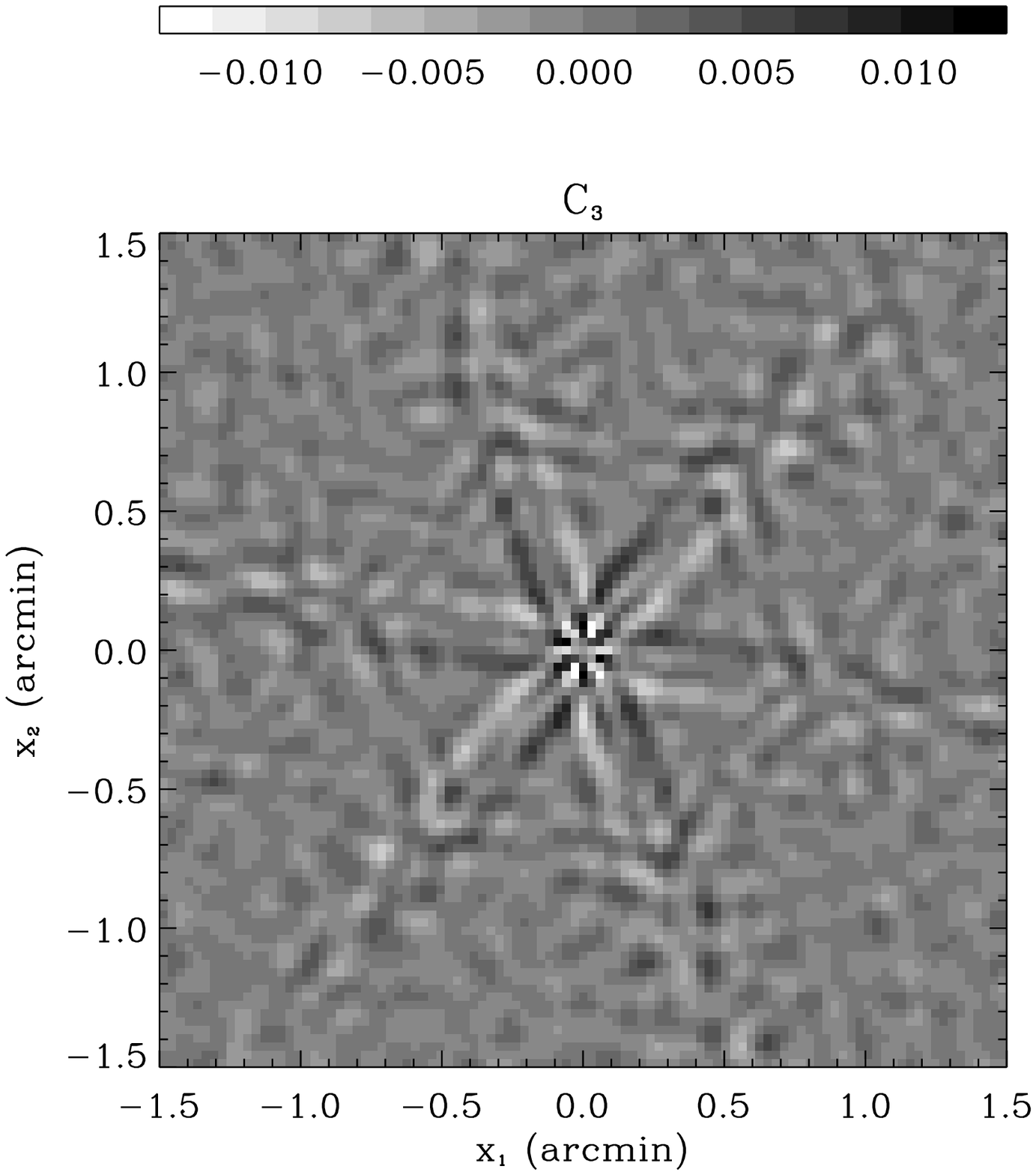}
\caption{[See caption above]}
\end{figure}

\section{Application to the FIRST Radio Survey}
\label{first}
In this section, we apply the general formalism developed above to
the specific case of the FIRST radio survey (\markcite{bec95}Becker et
al. 1995; \markcite{whi96}White et al. 1996).  The current version of
the FIRST survey contains about $4 \times 10^{5}$ sources, and covers
about $4.3 \times 10^{3}$ square degrees. The survey fields were
observed at 1.4 GHz with the VLA in the B configuration. The
$5\sigma$ detection limit of the survey is about 0.75 mJy. The pixel
size for the survey maps is 1.8$^{\prime \prime}$, while the restoring
beam is 5.4$^{\prime \prime}$ (FWHM). The restoring beam is thus
reasonably well sampled. This allows us to take the continuous limit
(Eq.~[\ref{eq:cts_limit}]) in our calculation. In addition, the
relatively high angular resolution allows an accurate measurement of
source morphologies. We are in the process of attempting to detect the
effect of weak lensing by large-scale structure with this unique
database (\markcite{kam98}Kamionkowski et al. 1998;
\markcite{ref98}Refregier et al. 1998).

Because of the observing time limitations for a survey of this
magnitude, the observations had to be performed in the snapshot mode
with integration times of 5 seconds.  The interferometric data (or
``UV'' data) for each snapshot pointing was then Fourier transformed
to construct a ``dirty'' image. The resulting pointing images were
CLEANed using algorithms adapted from the standard AIPS deconvolution
package. The final survey fields were produced by
a weighted sum of the grid pointing images. At a given point,
approximately 12 pointings contribute to the coadded field. Sources
were then detected and their shape parameters measured, using a
two-dimensional gaussian fit.

Even though the grid pointing strategy was designed to optimize the
signal-to-noise ratio of the coadded images, the final maps still
suffer from the limited UV-coverage of the snapshot pointings. As a
result, typical coadded maps contain small but noticeable ``stripes''
in the noise (see figure~\ref{fig:coadd}).  In other words, the noise
is correlated. As we discussed above, this affects the source
parameters derived by the fitting routine. In particular, the effect
of correlated noise is a potential systematic effect in a search for
correlations in the source ellipticities produced by weak lensing.

\subsection{Noise Auto-Correlation Function}
As a specific case, we considered the coadded field 07210+29486E in
the FIRST survey. This field covers a solid angle of $46.5^{\prime}
\times 34.5^{\prime}$, with a pixel size of 1.8$^{\prime
\prime}$. Figure~\ref{fig:coadd} shows a portion of this field with
the contrast enhanced to make the noise more apparent.  Bright radio
sources are apparent in the right hand corner.  In addition, the
stripe-like nature of the noise is visible. For this analysis, pixels
with intensities above .6 mJy beam$^{-1}$ (corresponding to about $4
\sigma$) were excised from the field. The excised regions were then
padded by 5.4$^{\prime \prime}$ to avoid contamination of the noise by
the source side wings. The residual noise has a standard deviation of
$\sigma_{N} \simeq .15$ mJy beam$^{-1}$. (Intensities are quoted in
mJy beam$^{-1}$, were the beam refers to the restoring beam with a
FWHM diameter of 5.4$^{\prime \prime}$.) The median noise standard
deviation for the FIRST survey is 0.14 mJy beam$^{-1}$, with 95\% of
the FIRST area having a noise standard deviation less than 0.17 mJy
beam$^{-1}$ (\markcite{whi96}White et al. 1996). This makes the
coadded field 07210+29486E somewhat noisy, yet representative of the
FIRST survey.

Figure~\ref{fig:eta} shows our measurement of the noise ACF,
$\eta({\bf x})$, for this field. For this measurement, the SNR for
$\eta(0)$ is 60 (see Eq.~[\ref{eq:eta_snr}]). For convenience,
$\eta({\bf x})$ was normalized to 1 at ${\bf x}=0$ in this figure. In
addition to a prominent central peak, the ACF shows marked radial
structures. These structures are reminiscent of the VLA antenna
pattern and characterize the ``stripes'' which are visible in the
noise.

Each grid pointing is characterized by a dirty beam. (It is the
Fourier transform of the UV coverage and is equal to the dirty image
produced by a point source at the phase tracking center). In general,
the dirty beam shape varies from one grid pointing to another due to
differing geometries, flagged antennas, etc.  As a result, coadded
fields, which are composed of several grid pointings, do not have a
well defined dirty beam. It is nevertheless instructive to
qualitatively compare our measurement of the noise ACF with a typical
grid pointing dirty beam. Figure~\ref{fig:dbeam} shows such a dirty
beam for the grid pointing 07000+50218. The qualitative resemblance
with $\eta({\bf x})$ is striking. This shows that a large fraction of
the noise is effectively convolved with the dirty beam.

\subsection{Ellipticity Correlation Functions}
Of practical interest for weak lensing measurements, is the effect of
the correlated noise on the ellipticity correlation functions.
Figure~\ref{fig:ci_th} shows the rotated correlation functions
$C_{1}({\bf x})$, $C_{2}({\bf x})$ and $C_{3}({\bf x})$
(Eq~[\ref{eq:wijrr}]) for the coadded field of figure~\ref{fig:coadd}.
A source size of $a=2.45^{\prime \prime}$ was chosen. This corresponds
to a FWHM diameter of $5.76^{\prime \prime}$, that is to a deconvolved
FWHM diameter of $2^{\prime \prime}$, for a restoring beam of
$5.4^{\prime \prime}$. The source signal-to-noise ratio was set to
SNR$_{s}=5$. These parameters correspond to both the detection and
resolution limit of the FIRST survey and can thus be considered as
``worse-case'' parameters for a weak lensing search.

The symmetric patterns of $\eta({\bf x})$ (figure~\ref{fig:eta}) are
also clearly apparent on figure~\ref{fig:ci_th}. However, the patterns
are different for each of the correlation functions.  The first
correlation function, $C_{1}({\bf x})$ exhibits a pronounced star-like
pattern with radial elongation on each arm. On the other hand,
$C_{2}({\bf x})$ is mostly characterized by a negative correlation
close to the center. Finally, $C_{3}({\bf x})$ exhibits star-like
patterns but of much smaller amplitude than that of $C_{1}({\bf x})$
and $C_{3}({\bf x})$. The last fact is expected for a non-parity
violating noise pattern.

\begin{figure}
\plotone{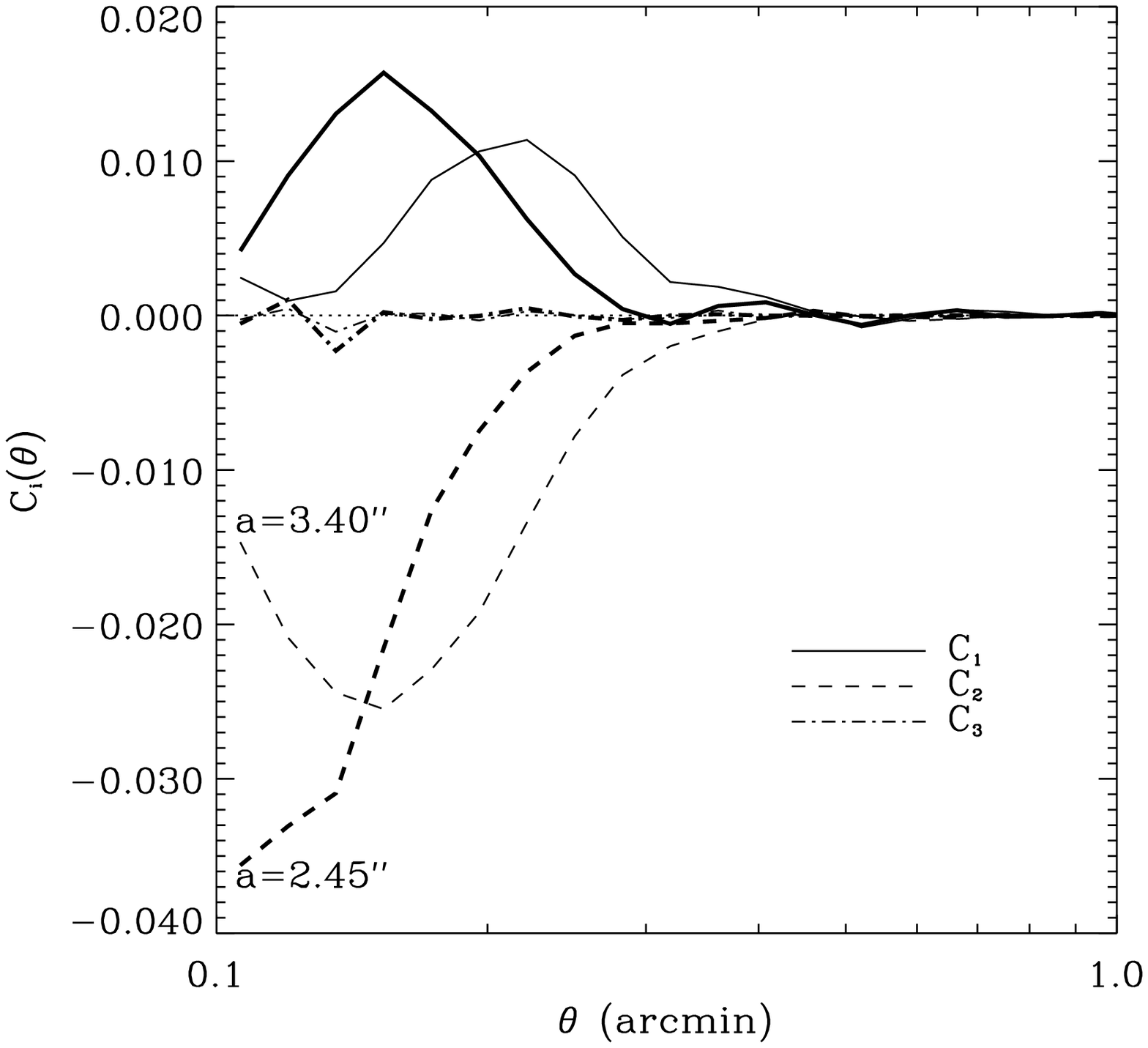}
\caption{Azimuthally-averaged ellipticity correlation functions for the
same coadded field. Each correlation functions, $C_{1}$, $C_{2}$,
and $C_{3}$, are plotted as a function of the separation radius
$\theta$. A source signal-to-noise ratio of SNR$_{s}=5$ was
assumed. The correlation functions scale as
$C_{i} \propto {\rm SNR}_{s}^{-2}$. The correlation functions are
plotted for a source size of $a=2.45^{\prime \prime}$ (thick lines) and
of $a=3.40^{\prime \prime}$ (thin lines).
\label{fig:ci_a}}
\end{figure}

Figure~\ref{fig:ci_a} shows the azimuthally averaged correlation
functions, $C_{1}(\theta)$, $C_{2}(\theta)$ and $C_{3}(\theta)$ as a
function of $\theta$. These are simply the radial profiles of the
rotated correlation functions of figure~\ref{fig:ci_th}.  They were
derived by Monte-Carlo averaging over $4 \times 10^{5}$ separation
vectors ${\bf x}$ picked randomly within a radius of $50^{\prime}$.
The thick lines correspond to a source size of $a=2.45^{\prime
\prime}$, while the thin lines corresponds to $a=3.40^{\prime
\prime}$. In both cases, the source signal-to-noise ratio was set to
SNR$_{s}=5$, as before. For $\theta \lesssim 0.4^{\prime}$,
$C_{1}(\theta)$ is positive while $C_{2}(\theta)$ is
negative and drops off faster. Note that both $C_{1}(\theta)$ and
$C_{2}(\theta)$ are significantly non-zero well beyond the source
radius $a$. This is of course due to the long-range features in
$\eta({\bf x})$ (figure~\ref{fig:eta}). At all angles, $C_{3}(\theta)$
remains smaller and oscillates around 0. As we remarked above, this is
expected for non-parity violating noise.

A comparison between the thick and thin lines in figure~\ref{fig:ci_a}
shows how the ellipticity correlation functions depend on the source
size. For larger source sizes, $C_{1}(\theta)$ and $C_{2}(\theta)$
keep the same qualitative behavior but have a lower maximum amplitude
and drop off more slowly. This is expected since larger source sizes
effectively smooth out the sharp features in the noise ACF
(figure~\ref{fig:eta}; see Eq.~[\ref{eq:wij_circ}]).
\label{a_behavior}

\begin{figure}
\plotone{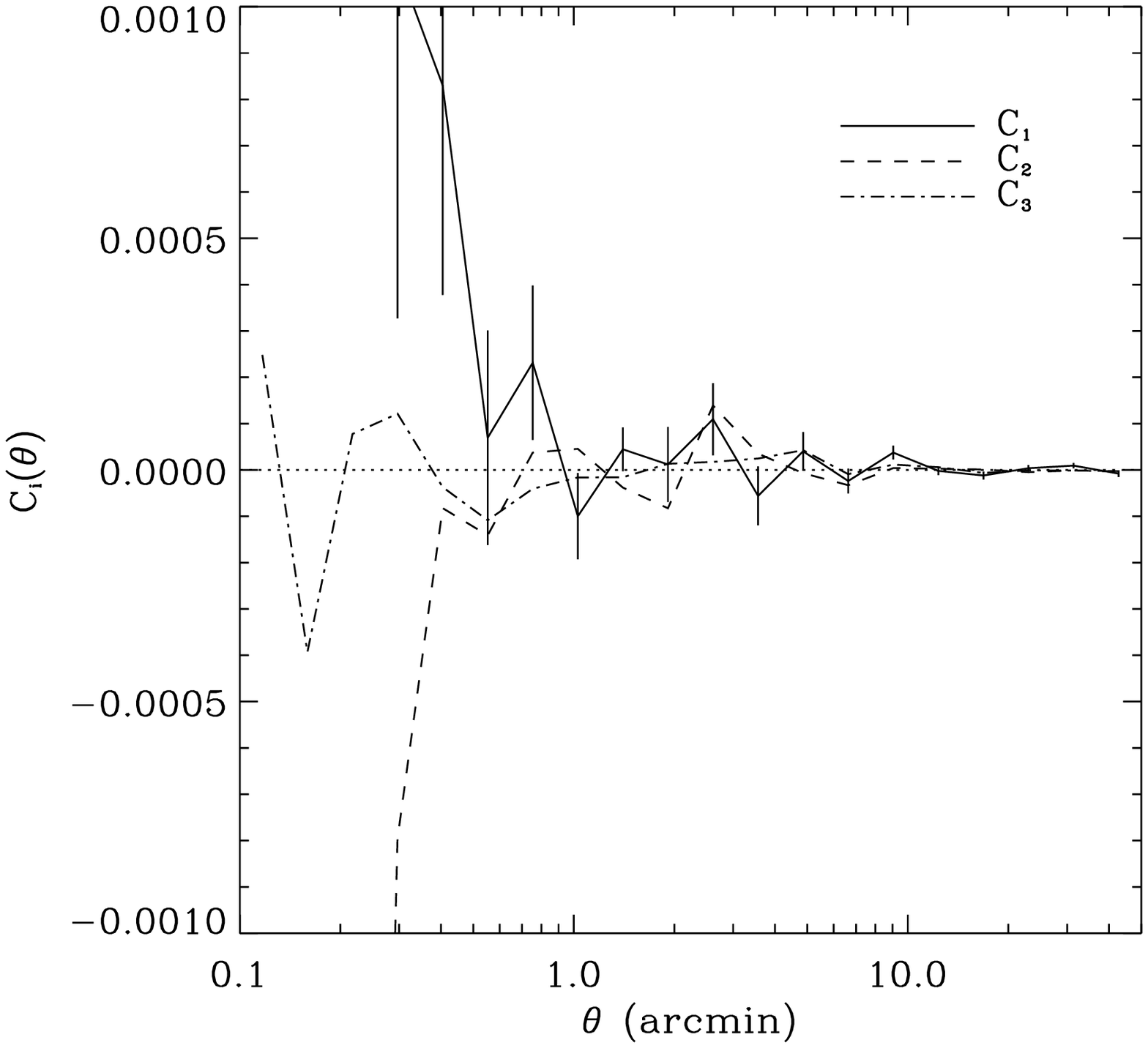}
\caption{Same as the previous figure but this time focusing
on the large-angle behavior of the correlation functions.
The correlation functions are only plotted for a source size of
$a=2.45^{\prime \prime}$. The error bars are the $1\sigma$
errors in the mean derived from the Monte-Carlo integration.
For clarity, they were only shown for $C_{1}(\theta)$.
The correlation functions are all consistent with 0 for
$\theta \gtrsim 1^{\prime}$.
\label{fig:ci_err}}
\end{figure}

Figure~\ref{fig:ci_err} shows the long range behavior of the
ellipticity correlation functions. For angles larger than
$\theta=10^{\prime}$, a mosaic of 4 fields contiguous with
07210+29486E was used to compute the noise ACF $\eta({\bf x})$.  The
error bars shown for $C_{1}(\theta)$ are the error in the mean derived
from the Monte-Carlo sampling described above. Since the separation
plane was undersampled, these error bars provide a good estimate of
the uncertainty in the correlation functions. As is apparent
on this figure, all three correlation functions are consistent
with 0 for $\theta \gtrsim 1^{\prime}$.

\subsection{Consequences for Weak Lensing Searches}
As was demonstrated above, correlated noise affects the ellipticity
correlation functions in the FIRST survey.  We now compare the
amplitude of this systematic effect with the signal expected for weak
lensing by large-scale structure.

\begin{figure}
\plotone{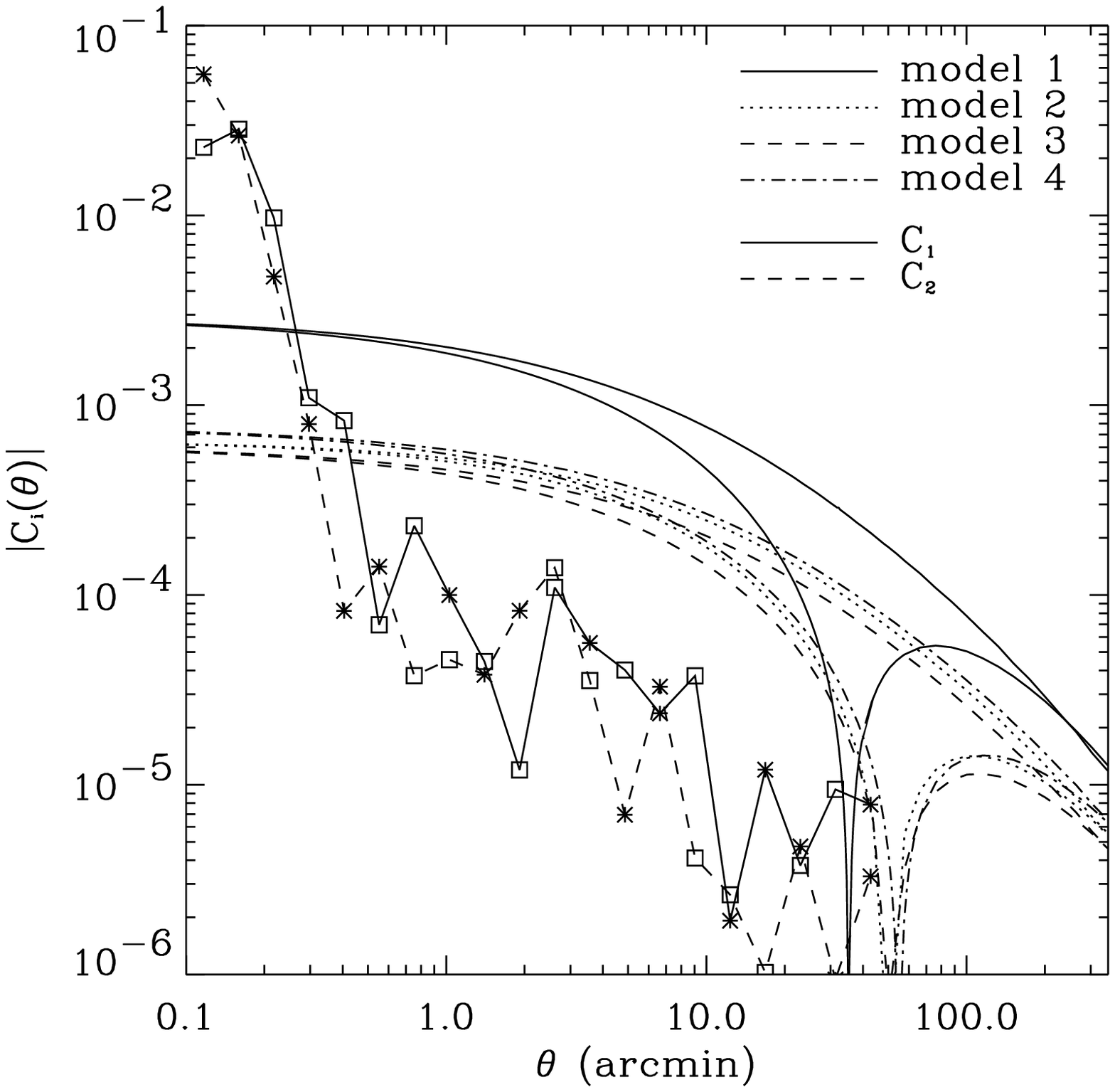}
\caption{Comparison with the weak lensing correlation functions expected
in CDM models. The smooth curves show
the ellipticity correlation functions for the four COBE-normalized CDM
models of Kamionkowski et al. (1998). The top and bottom curves
for each model correspond to $C_{1}$ and $C_{2}$, respectively.
The broken lines correspond to the ellipticity correlation functions
induced by the noise correlation, for sources with $a=2.45^{\prime \prime}$
and SNR$_{s}=5$. In all cases, the absolute value of the
correlation function is shown. To preserve a
sense of the noisy behavior of the noise-induced correlation functions,
positive and negative values are indicated by squares and stars,
respectively.
\label{fig:ci_theo}}
\end{figure}

Figure~\ref{fig:ci_theo} shows the noise-induced correlation
functions, $C_{1}(\theta)$ and $C_{2}(\theta)$ on a log-log scale.
The absolute value of the correlation functions are plotted, with
positive and negative values indicated by squares and stars,
respectively.  For comparison, the (absolute value of the) correlation
functions expected for weak lensing in CDM models are also
displayed. The four models correspond to the COBE-normalized CDM
models in table 2 and figure 4 in \markcite{kam98}Kamionkowski et
al. (1998).  Models 1-3 correspond to a standard, tilted, and lambda
CDM model, respectively. Model 4 corresponds to a standard CDM model
with a smaller Hubble constant than that of Model 1. In all models,
only linear evolution of density perturbations were considered.

As we noted above, noise correlation produces significant biases in
$C_{1}(\theta)$ and $C_{2}(\theta)$ for $\theta \lesssim 1^{\prime}$.
In this angular range, the noise-induced correlation functions exceed
those expected for weak lensing for all models. (The inclusion of
nonlinear evolution of density perturbations increases the estimates
of the weak lensing signal for $\theta \lesssim 10^{\prime}$ by only a
factor of a few and therefore does not significantly alter this
conclusion; see \markcite{jai97}Jain \& Seljak 1997.)

For angles $\theta \gtrsim 1^{\prime}$, we showed above that the
noise-induced correlation functions are consistent with zero, within
the Monte-Carlo uncertainties. This can be seen by the alternation of
squares and stars in figure~\ref{fig:ci_theo}. The noise-induced lines
in this figure should thus be interpreted as $1\sigma$ upper limits in
this angular range. (A larger number of Monte-Carlo simulations could
of course reduce this upper limit even further but is computationally
cumbersome.)  We notice that, for $1^{\prime} \lesssim \theta \lesssim
50^{\prime}$, these upper limits are almost one order of magnitude
smaller than model 3, the most pessimistic CDM model.

As we showed above, the noise-induced correlation functions depend
only moderately on the source size (see figure \ref{fig:ci_a}).  Most
of the sources used in our weak lensing searches have sizes close to
the resolution threshold ($a \simeq 2.45^{\prime \prime}$, or,
equivalently, FWHM deconvolved diameter of $2^{\prime\prime}$), and
hardly ever exceed $a\sim 5^{\prime \prime}$
(\markcite{ref98}Refregier et al. 1998). As a result, the
noise-induced correlation functions in such a survey will be close to
that shown on figure~\ref{fig:ci_theo}.

The Monte-Carlo realizations also allow us to compute the rms standard
deviation $\sigma[C_{i}(\theta)]$ of the ellipticity correlation
functions. For $10^{\prime} < \theta < 50^{\prime}$, we find
$\sigma[C_{1}(\theta)] \simeq \sigma[C_{2}(\theta)] \simeq 1.9 \times
10^{-3}$. This is almost two orders of magnitude below the standard
deviation produced by the intrinsic ellipticities of radio sources
($\sigma[C_{1}(\theta)] \simeq \sigma[C_{2}(\theta)] \approx
0.4^2=0.16$; see \markcite{ref98}Refregier et al. 1998). Thus,
statistical fluctuations in the noise-induced correlation functions
are negligible for $\theta \gtrsim 1^{\prime}$.

The estimation of the noise-induced correlation functions for $\theta
\gtrsim 50^{\prime}$ requires a large mosaic of coadded fields and is
therefore computationally cumbersome. However, we expect the
correlation functions to continue to be consistent with 0 at large
angles. In addition, a search for weak lensing will comprise an
average over the whole survey, as opposed to an average over a few
coadded fields in the present study. The dirty beam, and therefore
$\eta({\bf x})$, varies from one coadded field to the next. As a
result, we expect any long-range features in the noise ACF to average
out, leading to even smaller ellipticity correlations.

We conclude that, in the FIRST survey, noise-induced ellipticity
correlations dominate over the expected weak lensing signal for
$\theta \lesssim 0.5^{\prime}$, but are negligible for $\theta
\gtrsim 1^{\prime}$.

\section{Correlated Noise in Optical Images}
\label{optical}
While the effect is likely to be less severe than for intereferometric
radio surveys, noise correlation may also affect source shapes in
optical images. Most weak lensing studies are performed in the optical
band and are thus potentially sensitive to this effect. Noise
correlation in optical images can be produced by several effects.
First, noise correlation can arise from the convolution of background
light with the PSF. Because field distortions, diffraction spikes,
etc, can produce anisotropy of the PSF, this can induce ellipticity
correlations.  In addition, CCD readouts and charge-transfer
efficiency can potentially produce linear features in the noise
ACF. Finally, observing and image processing techniques such as
drift-scanning, shifting-and-adding, and ``drizzling''
(\markcite{fru97}Fruchter \& Hook 1997) can also produce features in
the ACF.

Measurements of the noise ACF in CCD images have recently been
performed by \markcite{van97}Van Waerbeke et al.  (1997), in the
context of a novel technique to measure weak lensing shear with the
noise ACF. In one of their fields, they find, at the center of the
noise ACF, evidence for a cross pattern which they tentatively
attribute to charge transfer efficiency and/or the shift-and-add
procedure.  Another application of their technique by
\markcite{sch97}Schneider et al. (1997) also indicates the presence of
similar instrumental effects at the center of the noise ACF for the
three CCD fields they considered. In one of the fields, the central
region of the noise ACF is clearly elongated along one of the CCD
axes. In another context, \markcite{vog97}Vogeley (1997) measured the
noise ACF of the Hubble Deep Field (HDF) in a search for fluctuations
in the Extragalactic Background Light. Preliminary tests indicate that
the extended wings of the PSF of the Hubble Space Telescope (HST)
produce as much as 10-40\% of this noise ACF signal. These wings are
probably due to internally scattered light. Other effects such as
flat-field errors could also contribute to the ACF but are likely to
be less important.

The effect of noise correlation should not be overlooked in weak
lensing searches with optical images. We have shown that the effect of
correlated noise is of second order in the inverse source SNR. One
must therefore be cautious in correcting for systematic shape
distortions using bright stars. While this technique, which is
standard in weak lensing studies in the optical, ensures proper
correction of the effect of the convolution of the PSF with the
sources, it would miss the effect of noise correlation. Indeed, for
stars with high SNR, the effect of the correlated noise is negligible,
while it could be significant for fainter galaxies used in weak
lensing searches.

It would thus be instructive to measure the noise ACF in optical
images with high SNR and to compute the ensuing ellipticity
correlations. This is one of the potential systematic effect in future
high precision searches of weak lensing by large-scale structure with
optical surveys (see \markcite{kai96}Kaiser 1996;
\markcite{ber97}Bernardeau et al. 1997).  In particular, the
drift-scanning technique involved in the processing of the Sloan
Digital Sky Survey could produce noise correlations which would need
to be corrected for in future searches of weak lensing with this
database (\markcite{ste95}Stebbins et al. 1995). The effect of noise
correlation could also be relevant for the recent detection of
galaxy-galaxy lensing in the HDF (\markcite{hud97}Hudson et al. 1997),
since this detection involves measurements of ellipticities at small
pair separations (a few arcsec) with the under-sampled PSF of the HST.

\section{Conclusions}
\label{conclusion}
We have studied the effect of noise correlation on the shape
parameters in two-dimensional images. The noise correlation is
conveniently described by the noise ACF which can easily be measured in
practice. We derived the magnitude of the effect for a general
two-dimensional least-square fit. The noise correlation can produce a
bias, and affect the variance and the covariance of source
parameters. In addition, it can produce systematic correlation in the
parameters of pairs of sources. We find that the effect is of second
order in the inverse SNR of the sources.

We applied these general results to the case of most practical
interest, namely a two-dimensional gaussian fit. We computed the
relevant matrices explicitly. In addition, we explicitly derived the
systematic bias produced by this effect on the ellipticity correlation
function of source pairs. This is particularly relevant for weak
lensing studies.

As a concrete example, we studied the effect of correlated noise on
the shape of sources in the FIRST radio survey. We measured the noise
ACF and found long range features which extend beyond the central
maximum. We computed the resulting systematic effect on the
ellipticity correlation function. We find that, in the FIRST survey,
noise-induced ellipticity correlations dominate over the expected weak
lensing signal for $\theta \lesssim 0.5^{\prime}$, but are
negligible for $\theta \gtrsim 1^{\prime}$.

We discussed the consequences of noise correlation for optical
surveys. In optical images, noise correlation can arise from various
effects such as the convolution of background light with the PSF, CCD
read outs, shift-and-add preprocessing, drift-scanning, etc.  Because
the effect is quadratic in the source SNR, the effect could be
overlooked if systematic distortions are monitored solely with bright
stars. The effect of noise correlation could thus be important for
searches of galaxy-galaxy lensing and of weak lensing by large-scale
structure in optical surveys.

\acknowledgments We first thank our collaborators in the FIRST
weak-lensing project for numerous discussions and exchanges.  They
include David Helfand, Marc Kamionkowski, Catherine Cress, Arif Babul,
Richard White, and Robert Becker. We are also indebted to Robert
Lupton, Michael Vogeley and Jim Gunn for useful discussions. AR was
supported by the NASA MAP/MIDEX program. STB was supported at Columbia
University with funds from an NSF Research Experience for
Undergraduates supplement to AST-94-19906 and NASA LTSA grant NAG
5-6035. This work was also supported by the NASA ATP grant NAG5-7154.

{} 

\end{document}